\documentclass{article}
\usepackage[utf8]{inputenc}
\usepackage{authblk}
\usepackage{setspace}
\usepackage[margin=1.25in]{geometry}
\usepackage{graphics}
\usepackage{graphicx}
\graphicspath{ {./figures/} }
\usepackage{booktabs}
\usepackage{array}
\usepackage{multirow}
\usepackage{subcaption}
\usepackage{amsmath}

\title{Feasibility of Li decorated Si$_{6}$C$_{14}$ and Si$_{8}$C$_{12}$ nanocages as promising hydrogen storage media: A computational study }

\author[1]{Ankita Jaiswal}
\author[2,3]{Rakesh K. Sahoo}
\author[1*]{Sridhar Sahu}

\affil[1]{Computational Materials Research Lab, Dept. of Physics, Indian Institute of Technology (Indian School of Mines) Dhanbad, Jharkhand, 826004, India}
\affil[2]{ Department of Physics, Karpagam Academy of Higher Education, Eachanari, Coimbatore, Tamil Nadu, 641021, India
}
\affil[3]{Department of Materials Science and Engineering, Technion-Israel Institute of Technology, Haifa, 3600003, Israel}
\date{}
\begin{document}

\maketitle

\begin{abstract}
	This article presents the reversible hydrogen storage capacities of Li decorated Si$_{6}$C$_{14}$ and Si$_{8}$C$_{12}$ using Density Functional Theory (DFT). The chemical stabilities of the designed Si$_{6}$C$_{14}$Li$_{6}$ and Si$_{8}$C$_{12}$Li$_{4}$ nanocages are investigated using HOMO-LUMO gaps and various global reactivity descriptors such as chemical hardness and electrophilicity index. Our study reveals that each Li atom decorated over the designed Si$_{6}$C$_{14}$ and Si$_{8}$C$_{12}$ nanocages can hold up to 5H$_{2}$ molecules with adsorption energy lying in the optimum range of 0.14-0.085 eV, thereby yielding to an overall gravimetric density of 13.8\% and 9.2\% respectively. The interaction between adsorbed H$_{2}$ molecules and the Li metal sites are found to occur via non-covalent and closed shell type of interaction.  The H$_{2}$ molecules are adsorbed in quasi-molecular fashion with elongated bond length. The molecular dynamics study reveals that most of the H$_{2}$ molecules get desorbed from the designed host nanocages at 300K without causing any significant structural changes, which confirmed their reversibility. When the adsorption and desorption conditions are set at 100K/60bar and 240K/1bar respectively, the practical storage gravimetric densities of Si$_{6}$C$_{14}$Li$_{6}$ and Si$_{8}$C$_{12}$Li$_{4}$ cages are estimated to be 13.73 wt\% and 9.08 wt\%, which are relatively high in comparison to the US-DOE target of 5.5 wt\% by the year 2020. Hence, the computationally designed Si$_{6}$C$_{14}$Li$_{6}$ and Si$_{8}$C$_{12}$Li$_{4}$ nanocages can be regarded as prospective systems for hydrogen storage applications.	

\end{abstract}

\section{Introduction}

The rapid depletion of fossil fuels and associated negative environmental repercussions, such as pollution and global warming, have heightened interest in developing alternative energy sources \cite{key-1,key-2}. Hydrogen is widely acknowledged as the most promising energy carrier because of its abundancy, maximum mass energy density and clean combustion \cite{key-3,key-4}. In comparison to gasoline, hydrogen produces at least three times as much chemical energy per unit mass. But from the standpoint of using hydrogen energy, its storage and production are the two significant challenges \cite{key-5}. Hydrogen can be produced artificially using environmental-friendly processes
like water electrolysis, photocatalysis and thermo-chemical cycles. The main constraint in the establishment and growth of a  hydrogen-centric economy lies in the development and practical use of hydrogen storage materials capable of storing hydrogen with a substantial gravimetric and volumetric density. These materials should also be feasible to implement under conditions close to normal atmospheric conditions in order to ensure optimal efficiency, safety, reliability, and cost-effectiveness in transportation systems \cite{key-6}. According to the target defined by the United States-Department of Energy (US-DoE), a hydrogen storage material can be deemed suitable for practical applications if it can achieve gravimetric density greater than 5.5 wt\% by the year 2020. This should be achievable within an optimal temperature range spanning from -40°C to 85°C, while also allowing for a maximum delivery pressure of 12 bar \cite{key-7}. The conventional hydrogen storage technologies currently in use, which include storing hydrogen gas compressed under high-pressure ($\sim$700-800 bar) or as a cryogenic liquid ($\sim$20 K), are associated with notable drawbacks like substantial energy losses, maintenance challenges, limited storage capacity, and safety hazards. This has incited scientific community to investigate material-based methods for storing hydrogen, wherein H$_{2}$ molecules are securely retained through the process of adsorption \cite{key-8}. Chemisorption of atomic hydrogens on a storage material with a high adsorption energy of (2-4 eV) has the potential to generate a high gravimetric density. The high desorption temperature, however, severely restricts its performance and practical application. In contrast, in the context of physisorption, H$_{2}$ molecules exhibit adsorption onto the host material with very low adsorption energy ($\sim$meV), enabling for simple desorption at lower temperature \cite{key-6}. However, research has shown that at near-ambient temperatures, a hydrogen storage material is considered to be ideal when it effectively stores hydrogen through weak non-covalent interactions in a quasi-molecular fashion, with an intermediate binding energy that lies between the chemisorbed and physisorbed states. According to recent studies, it has been determined that the ideal adsorption energy for H$_{2}$ adsorption in near-ambient applications should fall within the range of 0.1-0.2 eV \cite{key-9}. As a result, storage materials such as nanoclusters \cite{key-10,key-11}, fullerenes \cite{key-12,key-13}, and nanocages \cite{key-14,key-15} have kindled a lot of research interest recently. For material-based hydrogen storage systems, a variety of materials are already being investigated which include, metal hydrides \cite{key-16}, complex chemical hydrides \cite{key-17}, metal organic and inorganic frameworks \cite{key-18}, nanotubes \cite{key-19}, graphene sheet \cite{key-20} etc. However, due to slow kinetics, limited storage capacity, poor reversibility, high desorption temperatures, a lack of feasible realization, and inefficient application, most of the materials investigated have not been successfully put into practical applications with high efficiency. As a result, it has led researchers to either tailor and tune materials for efficient applications or to design, test and manufacture new materials that can fulfill the US-DoE criteria and technical challenges for hydrogen storage applications \cite{key-21}.

Extensive research has been conducted on carbon-based structures as prospective materials for hydrogen storage, owing to their advantageous characteristics such as low weight, significant porosity, and substantial surface-to-volume ratio.  However, the H$_{2}$ molecules adsorbed onto the surface of carbon-based structures exhibit a remarkably weak binding energy, leading to suboptimal adsorption \cite{key-22,key-23}. According to reports, the overall electronic properties of pristine nanostructures can be modified and rendered more favorable for H$_{2}$ adsorption through the use of different techniques, such as the substitution of foreign atoms like B, N, and Si \cite{key-24,key-25}. For instance, Si doped single-walled carbon nanotubes (SWNTs) and graphene are capable of adsorbing significant number H$_{2}$ molecules with enhanced binding energy than their pristine counterparts \cite{key-26}. According to recent research report Si doped C$_{60}$ fullerene can adsorb multiple molecular hydrogen resulting to a gravimetric density of 7.5wt\% \cite{key-27}. The presence of active metal centers, which produce a differential charge gradient on the surface and thus enhance the dipole interactions, has also been shown to enhance the adsorption of H$_{2}$ molecules in pure carbon-based structures through the decoration of metal atoms. This has been the subject of both experimental and theoretical research \cite{key-28,key-29}. A research report by Ammar et al. demonstrated that a single Ti atom functionalized over C$_{20}$ fullerene could adsorb 6H$_{2}$ molecules \cite{key-30}. However, investigations reveals that H$_{2}$ storage capacity of the transition metal functionalized cages is greatly reduced due to metal clustering. This occurs due to the exceptionally high cohesive energy of transition metals \cite{key-31,key-32}. In such instances, alkali metals such as Li are a superior alternative because of their relatively low cohesive energies, which reduces the potential of metal clustering \cite{key-33,key-34}. For example, Sun et al. studied that because of low cohesive energy, Li atoms decorated over C$_{60}$ fullerene does not undergo aggregation and can adsorb H$_{2}$ molecules with an average binding energy of about 0.075 eV/H$_{2}$ \cite{key-35}. Further research also shows that H$_{2}$ molecules adsorbed over Li$_{6}$C$_{60}$ desorb at low temperatures \cite{key-36}. Recent work by us has shown that pure C$_{20}$ fullerene when decorated with Li and Na can store 20H$_{2}$ molecules and thereby achieving a high gravimetric density of 13.08\% and 10.82\%, respectively \cite{key-37}. Metin et al. studied the encapsulation of molecular hydrogen inside C$_{20}$, Si- doped C$_{20}$ fullerenes and reported that C$_{15}$Si$_{5}$ was the most sensitive to the presence of H$_{2}$ molecules \cite{key-38}. W. Huang and his group have studied that chain terminated C$_{20}$ fullerene can adsorb multiple H$_{2}$ molecules with average adsorption energy lying in the range of 0.30 eV-1.18 eV \cite{key-39}. Further, our recent study demonstrates the high H$_{2}$ storage capacity of Li decorated Si$_{2}$C$_{18}$ and Si$_{4}$C$_{16}$ nanocages \cite{key-40}.

In this article, we present the viability of Li decorated Si$_{6}$C$_{14}$ and Si$_{8}$C$_{12}$ nanocages as potential hydrogen storage materials. The Si$_{6}$C$_{14}$ and Si$_{8}$C$_{12}$ nanocages chosen for the study have been extensively investigated in literature \cite{key-41} but have not yet been experimentally produced in the laboratory. However, based on the successful experimental synthesis of C$_{20}$ fullerene and Si-doped heterofullerenes, it is possible that the experimental synthesis of Si$_{6}$C$_{14}$ and Si$_{8}$C$_{12}$ nanocages will become viable in the near future.

\section{Theory and computation}

In electronic structure calculations, the frontier molecular orbital energies are vital inputs which are extensively used to estimate various reactivity descriptors like ionization potential (I), electron affinity (A), HOMO-LUMO gap (E$_{g}$) \cite{key-42}.

In conceptual density functional theory (CDFT) \cite{key-43}, electron donation or acceptance induces an appreciable change in energy, which is measured by the finite differences (FD) between the energies of the neutral system and its corresponding cationic and anionic systems with N, N-1, and N+1 electrons, respectively. Therefore, using Koopman's
theorem \cite{key-44}, I and A can be qualitatively expressed as

\begin{equation}
	I=E(N-1)-E(N)\approx-E_{HOMO}\label{eq:1}
\end{equation}

\begin{equation}
	A=E(N)-E(N+1)\approx-E_{LUMO}\label{eq:2}
\end{equation}
The structural stability and chemical reactivity of the studied complexes are described in terms of their binding energies (E$_{b}$), HOMO-LUMO gap (E$_{g}$) and global reactivity descriptors like hardness ($\eta$), electrophilicity ($\omega$) and electronegativity ($\chi$).

Chemical hardness ($\eta$) for a given system is expressed as

\begin{equation}
	\eta=\frac{I-A}{2}\approx\frac{E_{g}}{2}\label{eq:3}
\end{equation}

Therefore, we can conclude that chemically hard complexes will have higher E$_{g}$values and vice versa.
Similarly, for a given system, electronegativity ($\chi$) and eletrophilicity ($\omega$) can be expressed by the following relations

\begin{equation}
	\chi=\frac{I+A}{2}\label{eq:4}
\end{equation}

\begin{equation}
	\omega=\frac{\chi^{2}}{2\eta}\label{eq:5}
\end{equation}

The average binding energy ($E_{b}$) of the Li decorated over Si$_{6}$C$_{14}$ and Si$_{8}$C$_{12}$ nanocages are determined using the following relation

\begin{equation}
	E_{b}=\frac{E_{mLi+nanocage}-E_{nanocage}-mE_{Li}}{m}\label{eq:6}
\end{equation}
where $E_{mLi+nanocage}$, $E_{nanocage}$ and $E_{Li}$ are the total energy with m number of Li atoms adsorbed on heterofullerene, the global minimum energies of heterofullerene and an isolated Li atom, respectively.

The average adsorption energy $(E_{ads}$) of the H$_{2}$ molecules adsorbed over the nanocages is given by

\begin{equation}
	E_{ads}=\frac{\left[\left\{ E_{Host}+nE{}_{H_{2}}\right\} -E_{Complex}\right]}{n}\label{eq:7}
\end{equation}

In the above Eq.(\ref{eq:7}), $E_{H_{2}}$, $E_{Host}$ and $E_{Complex}$ are the minimum energies of the H$_{2}$ molecule adsorbed over the system, Li decorated Si$_{6}$C$_{14}$ and Si$_{8}$C$_{12}$ nanocages and their hydrogenated complexes respectively. The term $n$ signifies the number of H$_{2}$ molecules adsorbed over the designed nanocages.

The occupation number $(f)$ or the practical hydrogen storage capacity signifies the number of H$_{2}$ molecules remain adsorbed over the designed nanocages at a given temperature and pressure. Using grand canonical ensemble, it can be shown that the occupation number $(f)$ and grand canonical partition function ($Z$) are correlated as $f=k_{B}T\frac{\partial logZ}{\partial\mu}$ \cite{key-46}. Therefore, the average occupation number $(f)$ can be expressed as

\begin{equation}
	f=\frac{\sum_{n=0}^{n=n_{max}}ng_{n}exp(\frac{n(\mu-E_{ads})}{k_{B}T})}{\sum_{n=0}^{n=n_{max}}g_{n}exp(\frac{n(\mu-E_{ads})}{k_{B}T})}\label{eq:8}
\end{equation}
where, $k_{B}$, $T$ and $\mu$ indicate Boltzmann constant, temperature in Absolute scale and chemical potential of molecular hydrogen in gaseous phase respectively. The degeneracy term $g_{n}$ is taken as 1 after neglecting phonon contribution \cite{key-47}. Further, chemical potential ($\mu$) can be expressed in terms of other thermodynamic parameters as

\begin{equation}
	\mu=H^{0}(T)-H^{0}(0)-TS^{0}(T)+k_{B}Tln(\frac{P}{P_{0}})\label{eq:9}
\end{equation}

In the above Eq. (\ref{eq:9}) $H^{0}$ and $S^{0}$ denote the enthalpy and entropy of molecular hydrogen gas at $1bar$ pressure. The reference pressure $P_{0}$ is usually considered as $1bar$. The data required for the estimation of $\mu$ at various temperature and pressure are taken from the Ref. \cite{key-48}.

For a given hydrogen storage complex, the gravimetric density (wt\%) can be calculated using the following relation:

\begin{equation}
	H_{2}\left(wt\%\right)=\left[\frac{M_{H_{2}}}{\left(M_{H_{2}}+M_{Host}\right)}\right]\times100\label{eq:10}
\end{equation}

where $M_{H_{2}}$ and $M_{Host}$ are the total atomic weight of the adsorbed H$_{2}$ molecules and the bare complex respectively.

The density functional theory (DFT) based computations are performed with 6-311+G(d,p) basis set and Hybrid meta-GGA functional m06-2x functional. M06-2x functional consists of 54\% HF exchange and is nowadays widely used functional for the adsorption related studies involving non-covalent interactions in main group clusters \cite{key-51} All
these computation were carried out using Gaussian 09 software \cite{key-49}. The geometry optimizations of the pristine nanocages, Li functionalized complexes and their hydrogenated systems are done without any symmetry constraint. The density of states and partial density of states calculation are carried out using Gaussum software packages \cite{key-52}. Further, the bonding mechanism and the related interaction among different constituents of the complexes are understood in detail using quantum theory of atoms in molecules (QTAIM) \cite{key-53} using AIMALL software package \cite{key-54}.

\section{Results and discussions}

\subsection{Design of Li decorated Si$_{6}$C$_{14}$ and Si$_{8}$C$_{12}$ nanocages}

\begin{figure}[h!]
	\centering
	\includegraphics[scale=0.49]{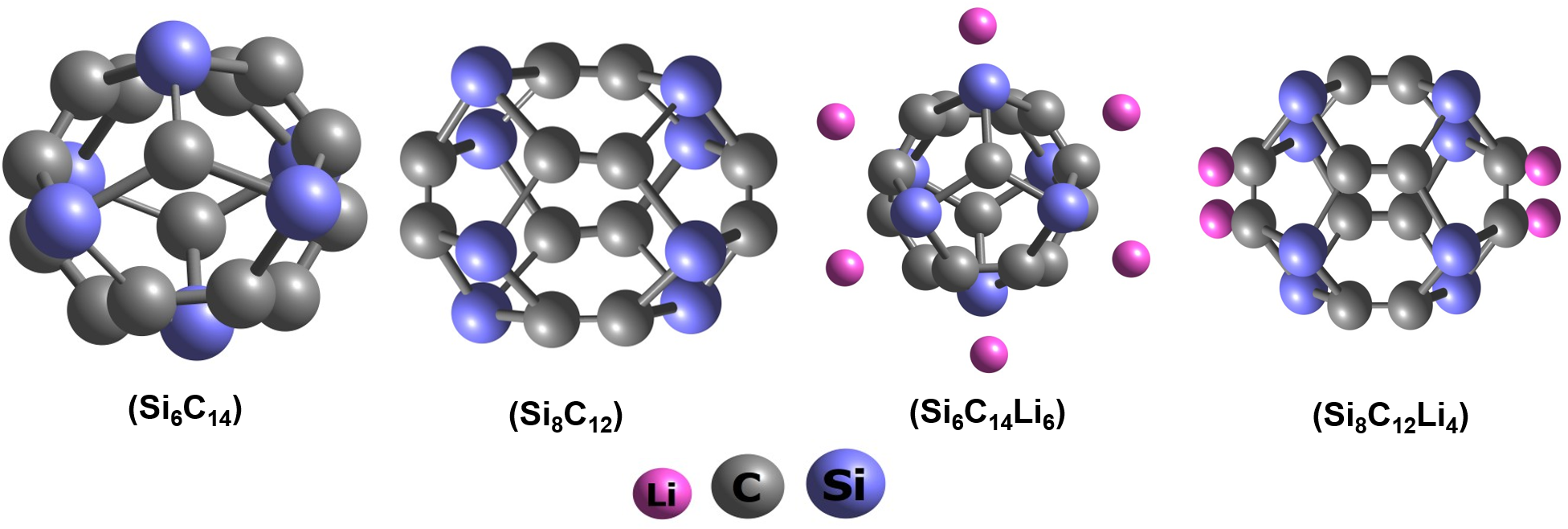}
	\caption{Optimized geometries of Si$_{6}$C$_{14}$, Si$_{8}$C$_{12}$, Si$_{6}$C$_{14}$Li$_{6}$ and Si$_{8}$C$_{12}$Li$_{4}$ nanocages.}
	\label{fig:fig1}
\end{figure}

Si$_{6}$C$_{14}$ and Si$_{8}$C$_{12}$ are Si substituted analogs of the smallest carbon fullerene C$_{20}$ which are theoretically well explored \cite{key-41}. The optimized geometry of Si$_{6}$C$_{14}$ nanocage depicted in Figure \ref{fig:fig1}, shows that it consists of twelve pentagons, These twelve pentagons forms two types of pentagonal rings;
in type1 six pentagonal rings consists of five carbon atoms and one Si atom whereas in type2- the other six pentagonal rings contains two Si atoms and four carbon atoms. However, Si$_{8}$C$_{14}$ nanocage as shown in Figure \ref{fig:fig1}, consists of four tetragonal rings with alternate C-Si bonds, four pentagonal rings containing one Si and
four carbon atoms and four hexagonal rings containing two Si atoms and four carbon atoms. To explore the potential of Si$_{6}$C$_{14}$ and Si$_{8}$C$_{12}$ nanocages as hydrogen storage candidates, it is necessary to achieve uniform functionalization of these cages with metal atoms. To computationally design the Li functionalization of Si$_{6}$C$_{14}$ and Si$_{8}$C$_{12}$ nanocages, it is necessary to identify the optimal site for Li binding, which is characterized by the lowest and maximum Li binding energies. This is modeled by minima hopping method \cite{key-55,key-56} in which a single Li atom is placed at each of the possible metal adsorption site and the
resultant geometry is then allowed to relax. In case of Si$_{6}$C$_{14}$ cage, there can be four possible ways of Li atom decoration- on the C-C bridge, on the C-Si bridge, atop of the pentagonal ring of type1 and atop of the pentagonal ring of type2. Using minima hopping method, it is found that the Li atom prefers to bind at the top of the pentagonal
ring of type1 with maximum binding energy of 3.94 eV. Therefore, six Li atoms are decorated at the top of each type1 pentagonal ring and the resultant structure is optimized to get our first computationally designed host structure Si$_{6}$C$_{14}$Li$_{6}$. Similar process is followed to decorate Li atoms over Si$_{8}$C$_{12}$ nanocage. As earlier revealed from the optimized geometry of Si$_{8}$C$_{12}$ cage that along with pentagonal rings and hexagonal rings, it also contains tetragonal rings with alternate C-Si bonds. Thus, in case of Si$_{8}$C$_{12}$ nanocage, there can be five possible ways of Li decoration- on the C-C bridge, on the C-Si bridge, on the top of the tetragonal ring, on the top of the pentagonal ring and at the top of the hexagonal ring. Li atom is regioselectively placed at each of the five possible sites of metal decoration and the resultant structures are allowed to relax. The global minimum energies of the optimized geometries of single Li atom decorated at different possible sites are then compared and it is found that Li atom prefers to bind at the top of the pentagonal sites with maximum binding energy of 3.05 eV. Therefore, four Li atoms are uniformly decorated at the top of each pentagons and the resultant structure is optimized to get our second designed host cage Si$_{8}$C$_{12}$Li$_{4}$. The functionalization of Li has no significant effect on the bond distances and the structural geometry of the hosts, as evident from Table \ref{tab:tab2}.

As the average binding energy of the Li atoms in both the designed Si$_{6}$C$_{14}$Li$_{6}$ and Si$_{8}$C$_{12}$Li$_{4}$
cages are found to be higher than the cohesive energy of Li in Li$_{2}$ dimer as well as in bulk phase, no  metal clustering can be expected. Minimum energy and frequency associated with all the possible Li decoration sites of Si$_{6}$C$_{14}$ and Si$_{8}$C$_{12}$ nanocages are tabulated in Table \ref{tab:tab1}.

\begin{table}[h!]
	\centering
	\caption{Energetically favourable site search for Li decoration over Si$_{6}$C$_{14}$
		and Si$_{8}$C$_{12}$ nanocages using minima hopping method.}
	\label{tab:tab1}
	\begin{tabular}{cccc}
		\toprule 
		Cluster & Possible Site & Minimum Energy (eV) & Min Frequency (cm$^{-1}$)\\
		\midrule 
		\multirow{4}{*}{Li@Si$_{6}$C$_{14}$} & C-C bridge & 0.06 & 70.1340\\
		& C-Si bridge & 0.02 & 101.6228\\
		& Pentagonal Ring1 & 0.0 & 71.2104\\
		& Pentagonal Ring2 & 0.25 & 94.4917\\
		\hline 
		\multirow{5}{*}{Li@Si$_{8}$C$_{12}$} & C-C bridge & 0.28 & 159.5063\\
		& C-Si bridge & 0.04 & 99.6283\\
		& Tetragonal Ring & 0.38 & 67.3742\\
		& Pentagonal Ring & 0.0 & 158.9867\\
		& Hexagonal Ring & 0.04 & 108.7690\\
		\bottomrule
	\end{tabular}
	
\end{table}

\subsubsection{Mechanism of Li functionalization}

\begin{figure}[h!]
	\centering
	\begin{subfigure}[h!]{0.485\textwidth}
		\centering
		\includegraphics[width=\textwidth]{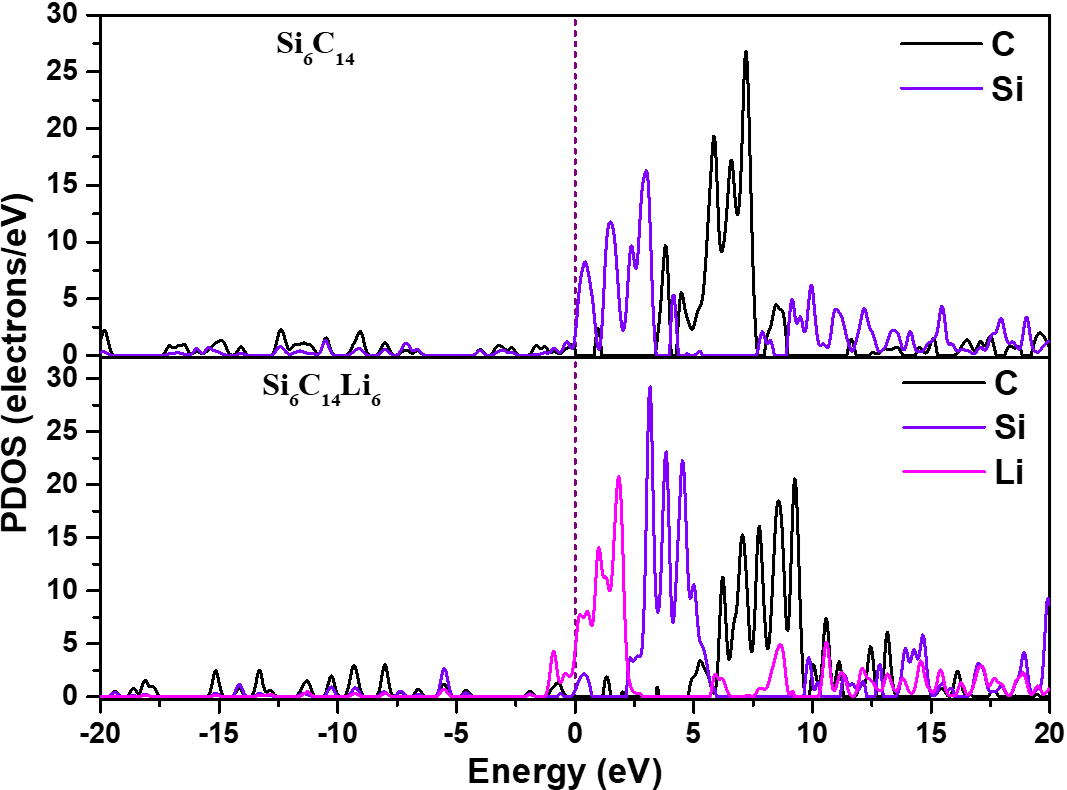}
		\caption{}
		\label{fig:fig2a}
	\end{subfigure}
	\hfill
	\begin{subfigure}[h!]{0.485\textwidth}
		\centering
		fig		\includegraphics[width=\textwidth]{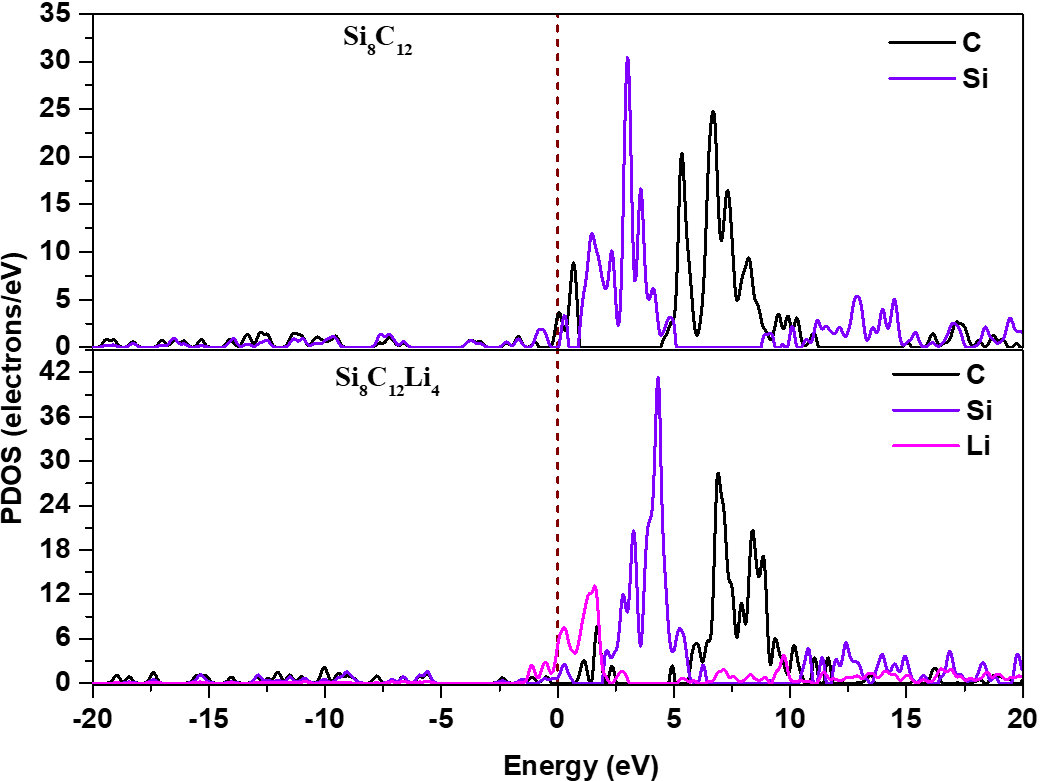}
		\caption{}
		\label{fig:fig2b}
	\end{subfigure}
	\caption{Variation in PDOS due to Li functionalization over (a) Si$_{6}$C$_{14}$ and (b) Si$_{8}$C$_{12}$ nanocages.}
	\label{fig:fig2}
\end{figure}

The binding mechanism of Li over Si$_{6}$C$_{14}$ and Si$_{8}$C$_{12}$ nanocages can be understood in terms of orbital interactions and charge transfer. This can be analysed by comparing the variation in the density of states of each constituent caused as a result of Li functionalization. The partial density of states plots of Si$_{6}$C$_{14}$ and Si$_{8}$C$_{12}$ nanocages and their corresponding Li functionalized derivatives are shown in Figure \ref{fig:fig2a} \& Figure \ref{fig:fig2b} respectively. Li functionalization over the nanocages is found to enhances the charge delocalization and can be easily understood by the appearence of new peaks in the density of states plot. No strong orbital overlap between Li, Si and C orbitals are observed which indicates there is no bond formation between Li and Si$_{6}$C$_{14}$ and Si$_{8}$C$_{12}$ nanocages. Due to Li functionalization, the density of states for Si enhances whereas for C the density of states reduces. This result is more prominent in Si$_{8}$C$_{12}$ and its Li decorated nanocages.

\begin{figure}
	\centering
	\includegraphics[scale=0.54]{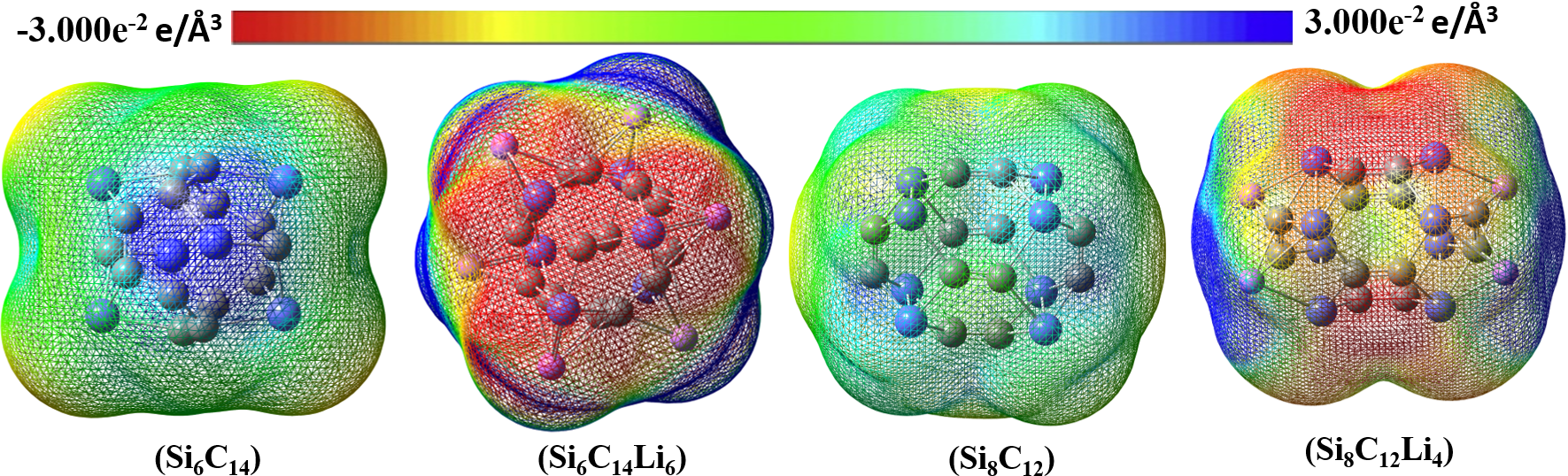}
	\caption{ESP maps of Si$_{6}$C$_{14}$, Si$_{6}$C$_{14}$Li$_{6}$, Si$_{8}$C$_{12}$ and Si$_{8}$C$_{12}$Li$_{4}$ nanocages.}
	\label{fig:fig3}
\end{figure}

To understand the binding mechanism of Li atom with Si$_{6}$C$_{14}$ and Si$_{8}$C$_{12}$ nanocages, electrostatic potential maps of nanocages before and after Li functionalization are plotted. The ESP maps of pristine Si$_{6}$C$_{14}$ and Si$_{8}$C$_{12}$ nanocages and their Li functionalized counterparts are depicted in terms of R-G-B colour codes which varies from red to blue as shown in Figure \ref{fig:fig3}. Red indicates the accumulation of electrons and, thus, a region of high electron density, whereas blue represents electron-deficient regions caused by electron depletion. Green colour lying in between them indicates region with moderate electron density.
Unfunctionalized Si$_{6}$C$_{14}$ and Si$_{8}$C$_{12}$ nanocages are initially greenish in colour, indicating their neutrality. However, Li functionalization causes an electron density shift from the decorated Li atoms to the Si$_{6}$C$_{14}$ and Si$_{8}$C$_{12}$ nanocages. As a result, the middle region turns crimson, while the outside regions
turn bluish. The difference in colour coding caused by Li functionalization implies that electron density has been redistributed, allowing the intended systems to adsorb additional H$_{2}$ molecules. To quantify the amount of charge transfer associated with electron density shift, Hirshfeld charge analysis has been carried out which indicates that
initially, the bare cage Si$_{6}$C$_{14}$ carried a small negative charge of approximately -0.0005 a.u. However, Li functionalization over the cage causes charge transfer from the decorated Li atoms to the cage. This charge transfer leads to charge gain of -2.025 a.u by the cage whereas each Li atoms decorated undergoes a charge loss of 0.34 a.u. In case of Si$_{8}$C$_{12}$ cage, the initial charge carried by the cage is measured to be -0.00016 a.u. However, upon
Li functionalization over the Si$_{8}$C$_{12}$ cage causes it to gain a charge of -1.35 a.u and the decorated Li atoms undergoes a charge loss of 0.34 a.u. Upon functionalization, Li atoms undergo charge loss which make them cationic in nature. As compared to neutral Li atom, the ionic Li atom is capable of adsorbing more H$_{2}$ molecules via charge polarization mechanism as proposed by Niu et al \cite{key-59}. This finding is in good consistent with the results of DOS analysis.

\subsubsection{Thermal Stability of the host cages}
\begin{figure}[h!]
	\centering
	\begin{subfigure}[h!]{0.485\textwidth}
		\centering
		\includegraphics[width=\textwidth]{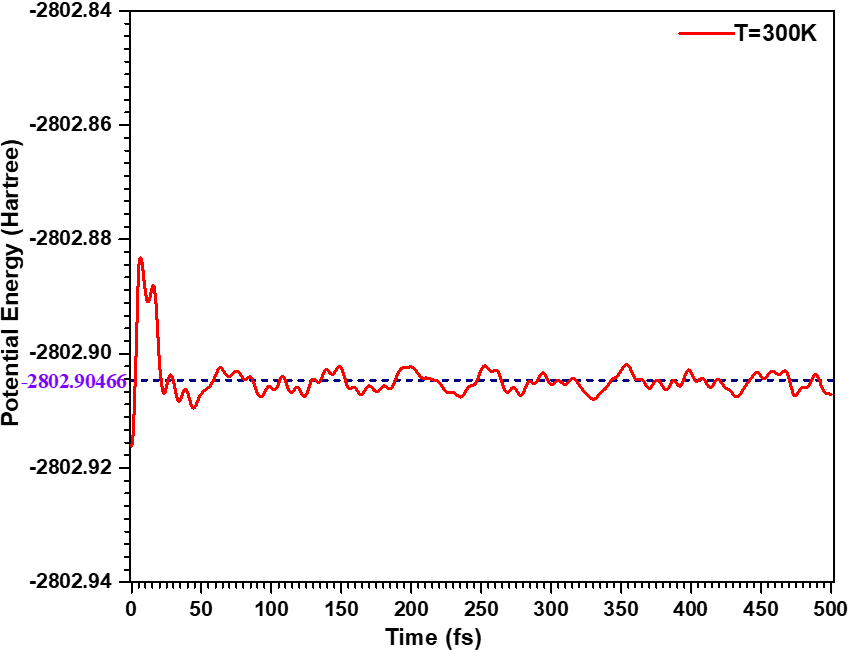}
		\caption{}
		\label{fig:fig4a}
	\end{subfigure}
	\hfill
	\begin{subfigure}[h!]{0.485\textwidth}
		\centering
		\includegraphics[width=\textwidth]{figures/fig4b.eps}
		\caption{}
		\label{fig:fig4b}
	\end{subfigure}
	\caption{Potential energy vs. time graph for (a) Si$_{6}$C$_{14}$Li$_{6}$
		and (b) Si$_{8}$C$_{12}$Li$_{4}$ nanocages.}
	\label{fig:fig4}
\end{figure}

To gain insight into the structural integrity and thermodynamic stability of the designed Si$_{6}$C$_{14}$Li$_{6}$ and Si$_{8}$C$_{12}$Li$_{4}$ nanocages, we carried out the ADMP-MD simulation at 300K for 500fs with time step of 0.5 fs. In both the designed nanocages, we found that the decorated Li atoms undergoes negligible changes and remain bonded to the surface of the designed cages which indicate the solidity and thermostability of the designed host cages. The variation of nearest neighbour bond distances C-Si and Si-Li for both the designed Si$_{6}$C$_{14}$Li$_{6}$ and Si$_{8}$C$_{12}$Li$_{4}$ nanocages are found to undergo a negligible change which further remarks the structural stability of the designed nanocages. Further, the potential energy vs. time graph shown in Figure \ref{fig:fig4} converges with time which further confirms the thermal stability of the designed nanocages.

\subsection{H$_{2}$ adsorption over Si$_{6}$C$_{14}$Li$_{6}$ and Si$_{8}$C$_{12}$Li$_{4}$ nanocages}
To determine the hydrogen storage capacities of the designed Si$_{6}$C$_{14}$Li$_{6}$ and Si$_{8}$C$_{12}$Li$_{6}$ cages, we have studied the sequential adsorption of H$_{2}$ molecules over them. We introduced the first sequence of H$_{2}$ molecules over Li decorated Si$_{6}$C$_{14}$ nanocage at a distance of about 2.1 Å and the structure is allowed
to relax. It is observed that the first sequence of H$_{2}$ molecules are adsorbed at a distance of 2.34 Å. The adsorption energy thus obtained is higher than the some of the previously reported results in similar systems. In a similar manner, more H$_{2}$ molecules are introduced in second, third, fourth and fifth sequential step. The optimized geometries of the sequentially hydrogen adsorbed Si$_{6}$C$_{14}$Li$_{6}$ and Si$_{8}$C$_{12}$Li$_{6}$ complexes are presented in Figure \ref{fig:fig6}. 

\begin{figure}[h!]
	\centering
	\includegraphics[scale=0.65]{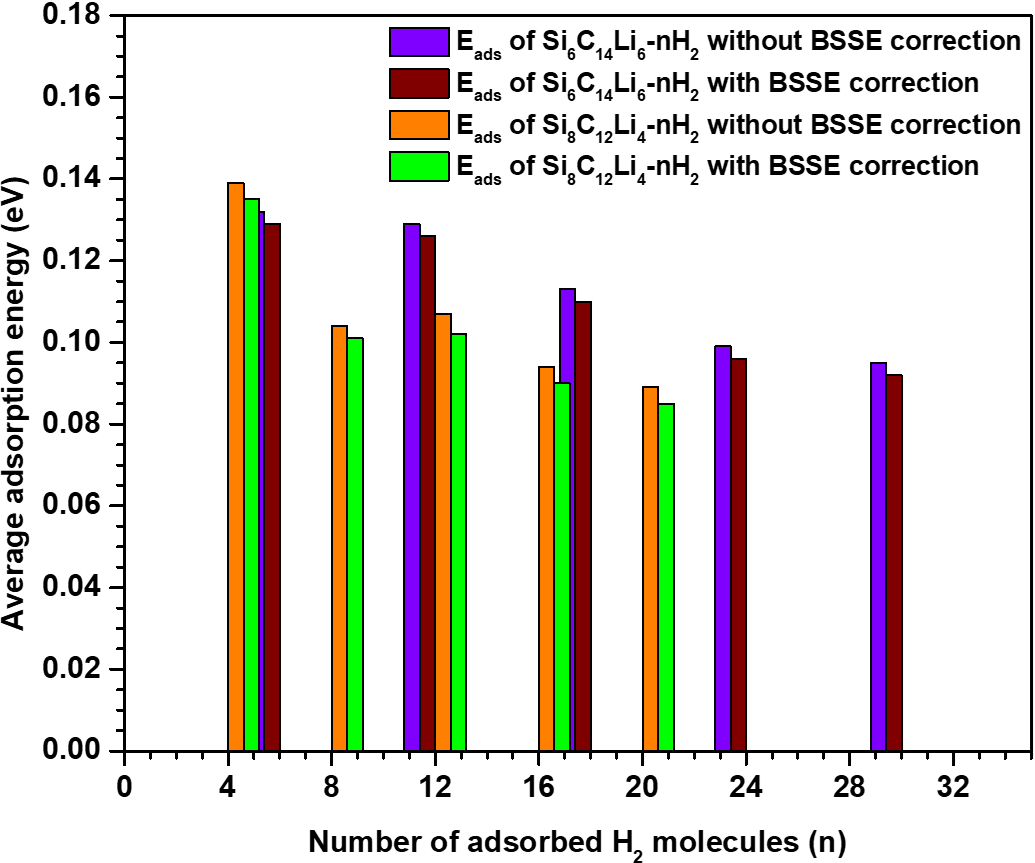}
	\caption{Variation of average adsorption energy with sequential H$_{2}$ adsorption
		over Si$_{6}$C$_{14}$Li$_{6}$ and Si$_{8}$C$_{12}$Li$_{4}$ nanocages.}
	\label{fig:fig5}
\end{figure}

\begin{figure}[h!]
	\centering
	\begin{subfigure}[h!]{0.98\textwidth}
		\centering
		\includegraphics[width=\textwidth]{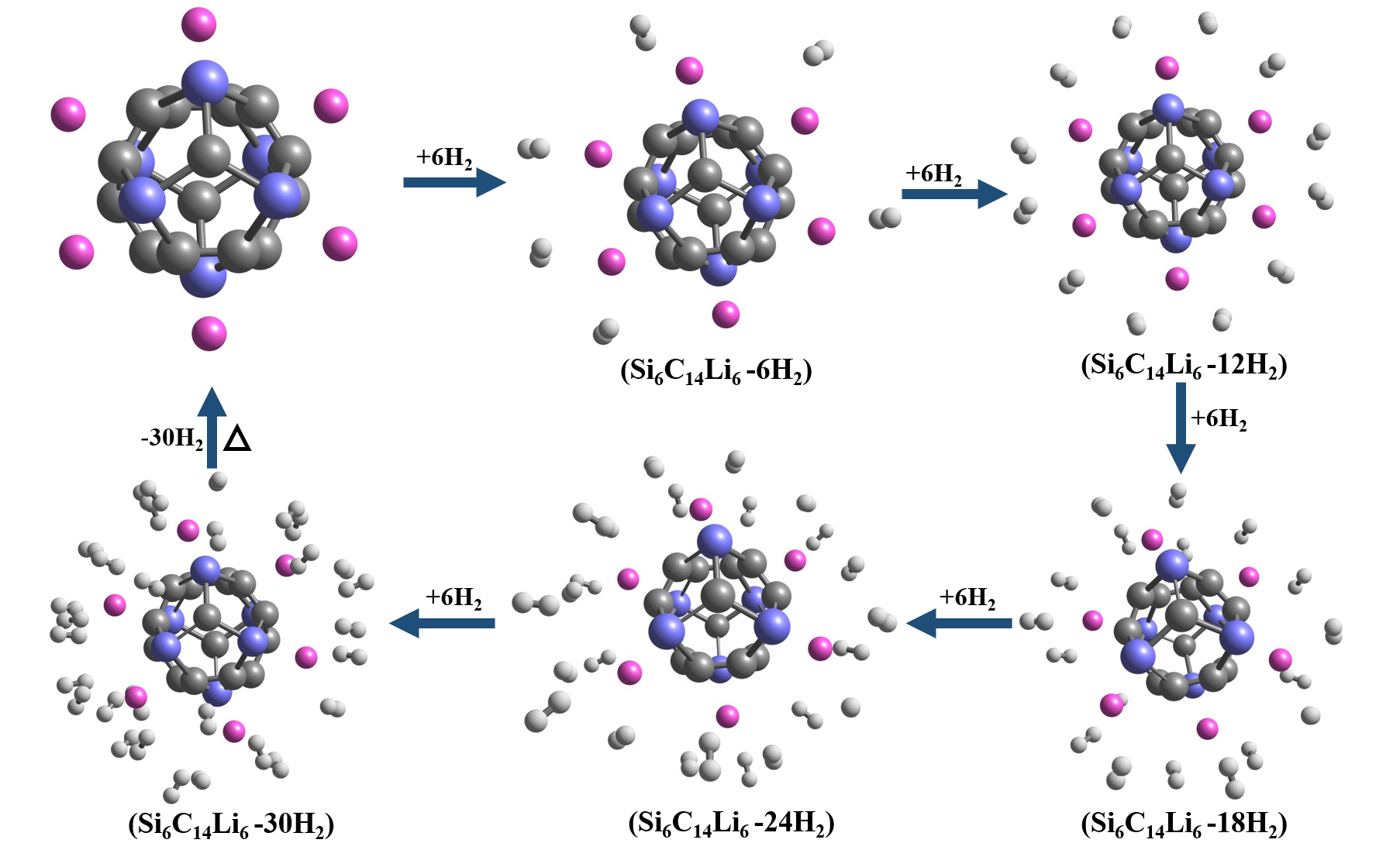}
		\caption{}
		\label{fig:fig6a}
	\end{subfigure}
	\quad
	\begin{subfigure}[h!]{0.95\textwidth}
		\centering
		\includegraphics[width=\textwidth]{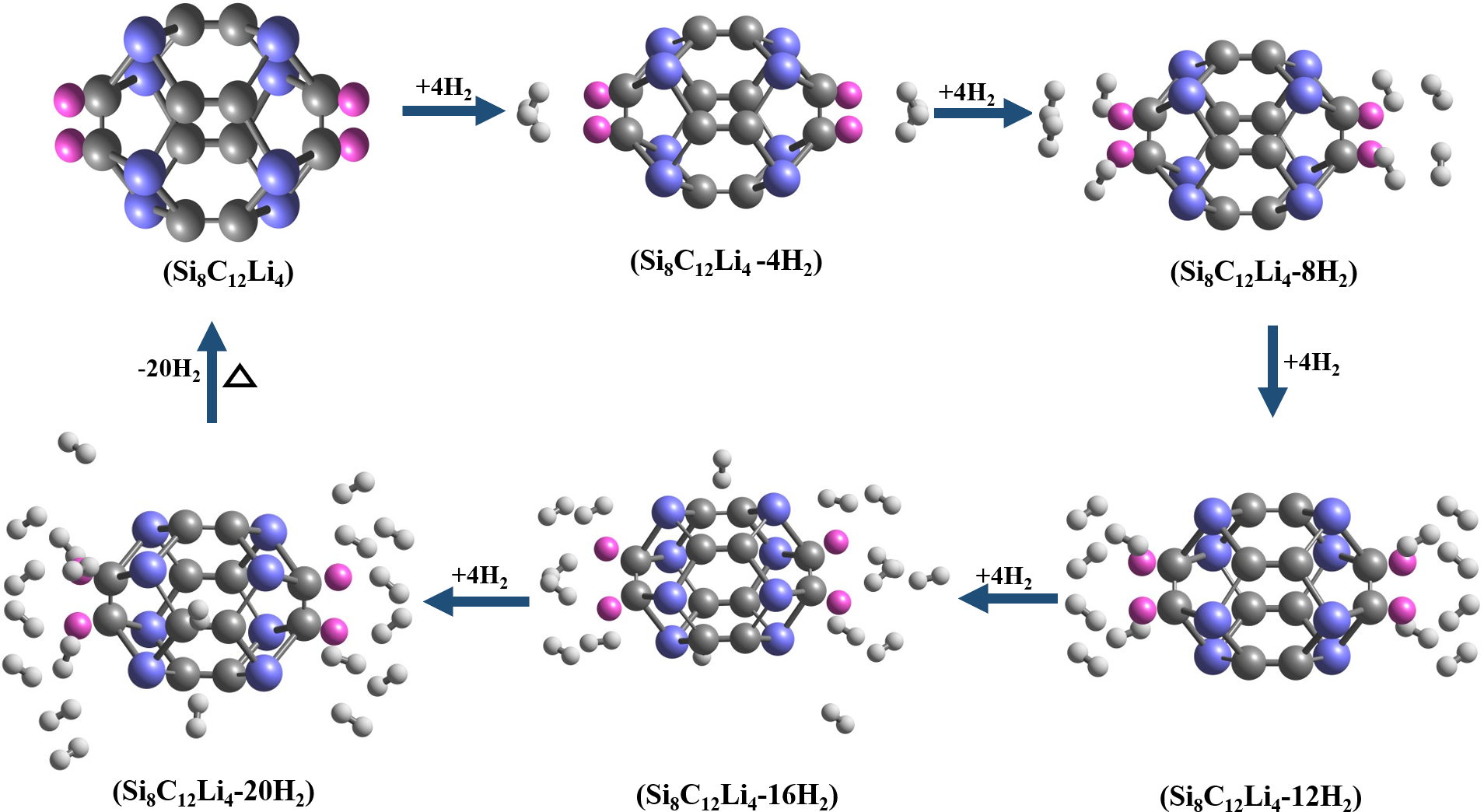}
		\caption{}
		\label{fig:fig6b}
	\end{subfigure}
	\caption{Successive adsorption of H$_{2}$ molecules over (a) Si$_{6}$C$_{14}$Li$_{6}$ and (b) Si$_{8}$C$_{12}$Li$_{4}$ nanocages.}
	\label{fig:fig6}
\end{figure}

For Si$_{6}$C$_{14}$Li$_{6}$ cage, the first H$_{2}$ molecule gets adsorb with an adsorption energy of 0.13 eV with an elongated H-H bond length of 0.75 Å. The average adsorption energy per H$_{2}$ molecule calculated using Eq. (\ref{eq:7}) is found to lie in the range of 0.13-0.095 eV. Further, the repulsion among the nearby H$_{2}$ molecules also causes adsorption at slightly larger distances. The average distance between the sorption centers and the adsorbed H$_{2}$ molecules are found to be in the range of 2.05-2.92 Å. For Si$_{8}$C$_{12}$Li$_{4}$, the first sequential adsorption of H$_{2}$ molecules occur with a binding energy of 0.14 eV at an average distance of 2.04 Å from the Li center. Each Li in Si$_{6}$C$_{14}$Li$_{6}$ and Si$_{8}$C$_{12}$Li$_{4}$ cage can adsorb up to 5H$_{2}$ molecules, thereby achieving a gravimetric density of 13.8\% and 9.2\% respectively. The variation of adsorption energy with sequential addition of H$_{2}$ molecules over the surface of Si$_{6}$C$_{14}$Li$_{6}$ and Si$_{8}$C$_{12}$Li$_{6}$ cages is depicted in Figure \ref{fig:fig5}. Successive adsorption shows that with the increase in H$_{2}$ adsorption over both the cages, the H$_{2}$ molecules are adsorbed slightly at larger distances and with lower binding energy because of space constraint and steric hindrance. In addition, the BSSE corrected adsorption energies are found little bit lower. However, the change in adsorption energies due to BSSE consideration is not very significant, which is consistent with our earlier report \cite{key-13}. The variation of various geometrical parameters as a result of Li functionalization and sequential H$_{2}$ adsorption over the cages are given in Table \ref{tab:tab2}. In both the host cages, the range of adsorption energy and the distances between the Li sorption center and the adsorbed H$_{2}$ indicates quasi-molecular adsorption.
\begin{table}[h!]
	\centering	
	\caption{Average bond lengths of C-C, C-Si, H-H, distance of Li from the center
		of the ring (Li-ring) and from the adsorbed H$_{2}$
		molecules (Li-H$_{2}$) in $\textrm{Å}$.}
	\label{tab:tab2}
	\begin{tabular}{cccccc}
		\toprule 
		Cluster & C-C & C-Si & Li-ring & Li-H$_{2}$ & H-H\\
		\midrule
		Si$_{6}$C$_{14}$ & 1.418 & 1.869 & - & - & -\\
		Si$_{6}$C$_{14}$Li$_{6}$ & 1.437 & 1.839 & 1.790 & - & -\\
		Si$_{6}$C$_{14}$Li$_{6}$-6H$_{2}$ & 1.436 & 1.843 & 1.821 & 2.335 & 0.747\\
		Si$_{6}$C$_{14}$Li$_{6}$-12H$_{2}$ & 1.437 & 1.965 & 1.811 & 2.052 & 0.748\\
		Si$_{6}$C$_{14}$Li$_{6}$-18H$_{2}$ & 1.436 & 1.965 & 1.835 & 2.399 & 0.747\\
		Si$_{6}$C$_{14}$Li$_{6}$-24H$_{2}$ & 1.436 & 1.964 & 1.860 & 2.661 & 0.746\\
		Si$_{6}$C$_{14}$Li$_{6}$-30H$_{2}$ & 1.434 & 1.962 & 1.970 & 2.922 & 0.746\\
		Si$_{8}$C$_{12}$ & 1.293 & 1.979 & - & - & -\\
		Si$_{8}$C$_{12}$Li$_{4}$ & 1.325 & 1.952 & 1.960 & - & -\\
		Si$_{8}$C$_{12}$Li$_{4}$-4H$_{2}$ & 1.325 & 1.952 & 1.967 & 2.036 & 0.747\\
		Si$_{8}$C$_{12}$Li$_{4}$-8H$_{2}$ & 1.324 & 1.954 & 1.977 & 2.176 & 0.745\\
		Si$_{8}$C$_{12}$Li$_{4}$-12H$_{2}$ & 1.327 & 1.952 & 2.039 & 2.199 & 0.746\\
		Si$_{8}$C$_{12}$Li$_{4}$-16H$_{2}$ & 1.327 & 1.953 & 2.036 & 2.660 & 0.746\\
		Si$_{8}$C$_{12}$Li$_{4}$-20H$_{2}$ & 1.327 & 1.954 & 2.043 & 2.952 & 0.745\\
		\bottomrule
	\end{tabular}
\end{table}

The stability of the hydrogen adsorbed systems are analyzed using HOMO-LUMO gap (E$_{g}$), global reactivity descriptors like hardness ($\eta$) and electrophilicity index ($\omega$). A larger E$_{g}$ signifies that a greater amount of energy will be needed by an electron to move from HOMO to LUMO level. The HOMO-LUMO gap, hardness and electrophilicity index of the studied systems are tabulated in Table \ref{tab:tab3}. From the tabulated data, we can observe negligible change in HOMO-LUMO gap and chemical hardness values increases due to Li functionalization and sequential H$_{2}$ adsorption. As the sequential $H_{2}$ adsorption process approaches saturation, there is an evident decrease in the values of Eg. We can therefore argue that the maximum hardness concept is being obeyed to some extent \cite{key-60}. However, the electrophilicity values follow a decreasing trend, demonstrating that the minimal electrophilicity principle \cite{key-61} is adhered to as a result of Li functionalization and successive $H_{2}$ adsorption across Si$_{6}$C$_{14}$Li$_{6}$ and Si$_{8}$C$_{12}$Li$_{4}$ nanocages. Thus, following the principle of maximum hardness and minimal electrophilicity ensures the stability of the studied systems.

\begin{table}[h!]
	\centering
	\caption{E$_{g}$, $\eta$ and $\omega$ of the studied systems in eV.}
	\label{tab:tab3}
	\begin{tabular}{cccc}
		\toprule 
		Cluster & E$_{g}$ & $\eta$ & $\omega$\\
		\midrule 
		Si$_{6}$C$_{14}$ & 2.66  & 1.328  & 4.142 \\
		Si$_{6}$C$_{14}$Li$_{6}$ & 2.69  & 1.346  & 1.942 \\
		Si$_{6}$C$_{14}$Li$_{6}$-6H$_{2}$ & 2.73  & 1.364  & 1.773 \\
		Si$_{6}$C$_{14}$Li$_{6}$-12H$_{2}$ & 2.77  & 1.383 & 1.627\\
		Si$_{6}$C$_{14}$Li$_{6}$-18H$_{2}$ & 2.74 & 1.371 & 1.622\\
		Si$_{6}$C$_{14}$Li$_{6}$-24H$_{2}$ & 2.71 & 1.357 & 1.626\\
		Si$_{6}$C$_{14}$Li$_{6}$-30H$_{2}$ & 2.62 & 1.312 & 1.571\\
		Si$_{8}$C$_{12}$ & 2.78 & 1.388 & 3.929\\
		Si$_{8}$C$_{12}$Li$_{4}$ & 3.21 & 1.604 & 2.417\\
		Si$_{8}$C$_{12}$Li$_{4}$-4H$_{2}$ & 3.2 & 1.601 & 2.325\\
		Si$_{8}$C$_{12}$Li$_{4}$-8H$_{2}$ & 3.15 & 1.577 & 2.219\\
		Si$_{8}$C$_{12}$Li$_{4}$-12H$_{2}$ & 3.13 & 1.564 & 2.162\\
		Si$_{8}$C$_{12}$Li$_{4}$-16H$_{2}$ & 3.14 & 1.570 & 2.188\\
		Si$_{8}$C$_{12}$Li$_{4}$-20H$_{2}$ & 3.17 & 1.585 & 2.186\\
		\bottomrule
	\end{tabular}
\end{table}

\subsubsection{ESP and Hirshfeld charge analysis}

\begin{figure}[h!]
	\centering
	\includegraphics[scale=0.49]{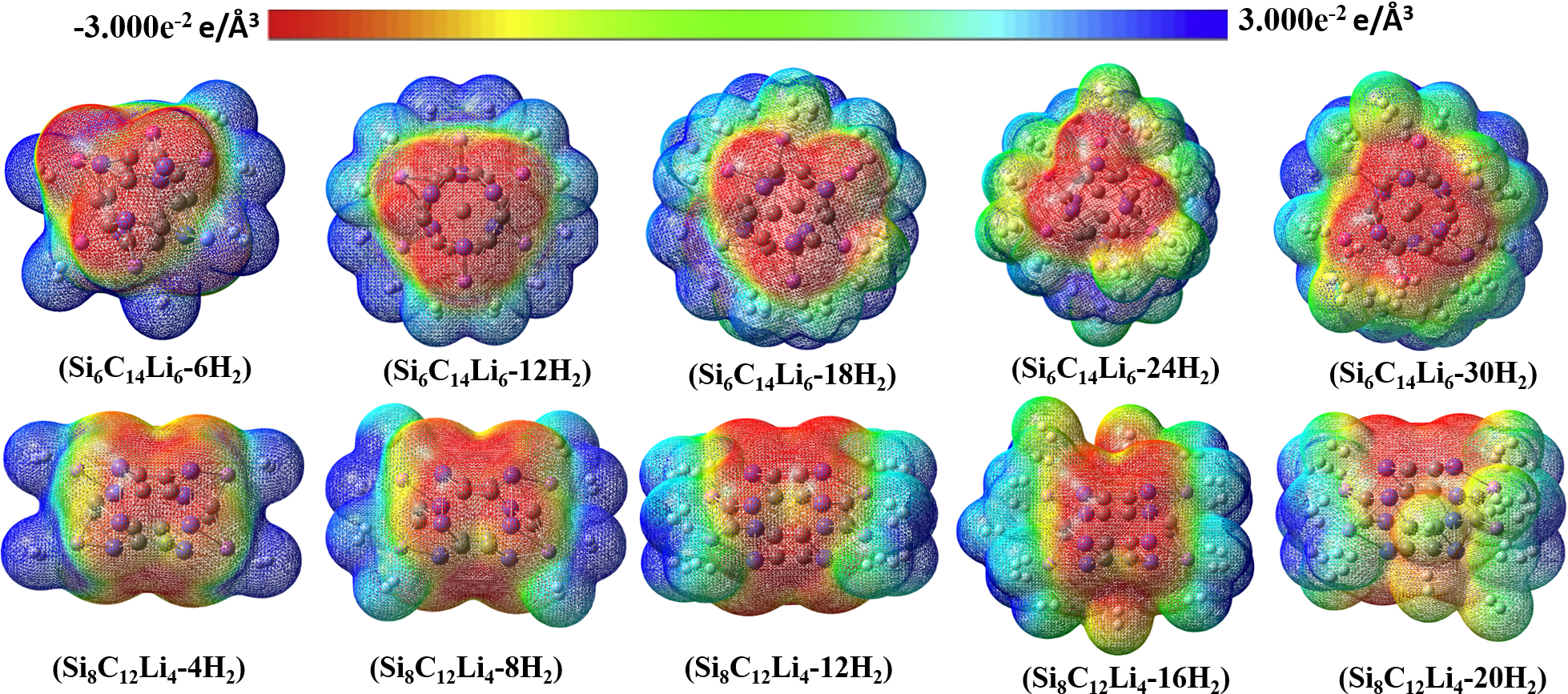}
	\caption{ESP maps of sequential hydrogen adsorbed Si$_{6}$C$_{14}$Li$_{6}$
		and Si$_{8}$C$_{12}$Li$_{4}$ systems.}
	\label{fig:fig7}
\end{figure}

Surface mapping of electrostatic potential over total electron density is used to study the electron density distribution that occurs as a result of sequential H$_{2}$ adsorption. As discussed earlier, in ESP maps, the change in electron density can be visualized in terms of changing colour codes. The ESP maps of the successive hydrogenated systems of Si$_{6}$C$_{14}$Li$_{6}$ and Si$_{8}$C$_{12}$Li$_{4}$ are depicted in Figure \ref{fig:fig7}. 
Obviously successive adsorption of H$_{2}$ molecules over Si$_{6}$C$_{14}$Li$_{6}$ and Si$_{8}$C$_{12}$Li$_{4}$ causes redistribution of the total electron density over the whole system, and it can be seen that the central portion of Si$_{6}$C$_{14}$ and Si$_{8}$C$_{12}$ nanocages become more reddish (gain of electron density) after H$_{2}$ adsorption. On the other hand, the color over Li atoms appears to fade after H$_{2}$ adsorption. The color shifts from red to yellowish-green, indicating that the electron density over these atoms drops, resulting in positive polarity. When the first sequential hydrogen adsorbed systems are compared with the saturated ones, we observe that the colour over H$_{2}$ molecules changes from blue to green, indicating transfer of electron density from Li atoms to H$_{2}$ molecules and thereby triggering the adsorption process. The shifting of electron density in ESP maps clearly indicates that charge transfer may be accompanied in the adsorption process. The variation of electron density as a result of H$_{2}$ adsorption is associated with charge transfer. In order to understand the charge transfer caused due to H$_{2}$ adsorption process over the Si$_{6}$C$_{14}$Li$_{6}$ and Si$_{8}$C$_{12}$Li$_{4}$ nanocages, we calculated average Hirshfeld charges over the adsorbed H$_{2}$ molecules, decorated Li atoms and the pristine Si$_{6}$C$_{14}$ and Si$_{8}$C$_{12}$ nanocages. Initially, the Hirshfeld charge over the Si$_{6}$C$_{14}$Li$_{6}$ heterofullerene is found to be 0.0003 a.u. But introduction of first sequence of H$_{2}$ molecules over the Li centers of Si$_{6}$C$_{14}$ cage, the total Hirshfeld charge on Si$_{6}$C$_{14}$Li$_{6}$ becomes -0.51 a.u which clearly indicates accumulation of electron density over the host cage and is marked by the reddish colour in Figure \ref{fig:fig7}. As more and more number of H$_{2}$ molecules are adsorbed over the Si$_{6}$C$_{14}$Li$_{6}$ and Si$_{8}$C$_{12}$Li$_{4}$, the Hirshfeld charge over Li goes on decreasing. Before hydrogenation, the Hirshfeld charge of each Li functionalized over Si$_{6}$C$_{14}$ heterofullerene carried is 0.337 a.u. However, in saturated hydrogen adsorbed system, the Hirshfeld charge of Li has significantly decreased to 0.181 a.u. which is about 53.7\% of the initial charge. Further, for the sequential H$_{2}$ adsorbed Si$_{6}$C$_{14}$Li$_{6}$ nanocages, the Hirsheld charge over the heterofullerenes decreases from -2.025 a.u to -1.865 a.u. Therefore, till H$_{2}$ adsorption reaches saturation, the Si$_{6}$C$_{14}$Li$_{6}$ loses about 16\% of charge. In addition, the average Hirshfeld charge over the adsorbed H$_{2}$ molecules goes on slight decreasing due to polarization as a result of sequential adsorption. Similar change is also observed in Si$_{8}$C$_{12}$Li$_{4}$-nH$_{2}$ complexes. However, Li and H continues to loose charge upon sequential H$_{2}$ adsorption. For example, successive hydrogen adsorption over Si$_{8}$C$_{12}$Li$_{4}$ causes total Hirshfeld charge of Li decreasing from 1.35-0.78 a.u which is about 42.2\% of the initial charge. Likewise, with successive adsorption, the average Hirshfeld charge of the adsorbed H$_{2}$ decreases drastically because of polarization. Therefore Hirshfeld charge analysis gives a qualitative as well as quantitative insight how Li atom donates charge to the Si$_{6}$C$_{14}$(Si$_{8}$C$_{12}$) cages, resulting to the formation of a dipole. The dipole formed because of polarization generates electrostatic field that helps to adsorb H$_{2}$ molecules through polarization which is also noticed form the above ESP studies. 

\subsubsection{PDOS analysis}

\begin{figure}[h!]
	\centering
	\begin{subfigure}[h!]{0.99\textwidth}
		\centering
		\includegraphics[width=\textwidth]{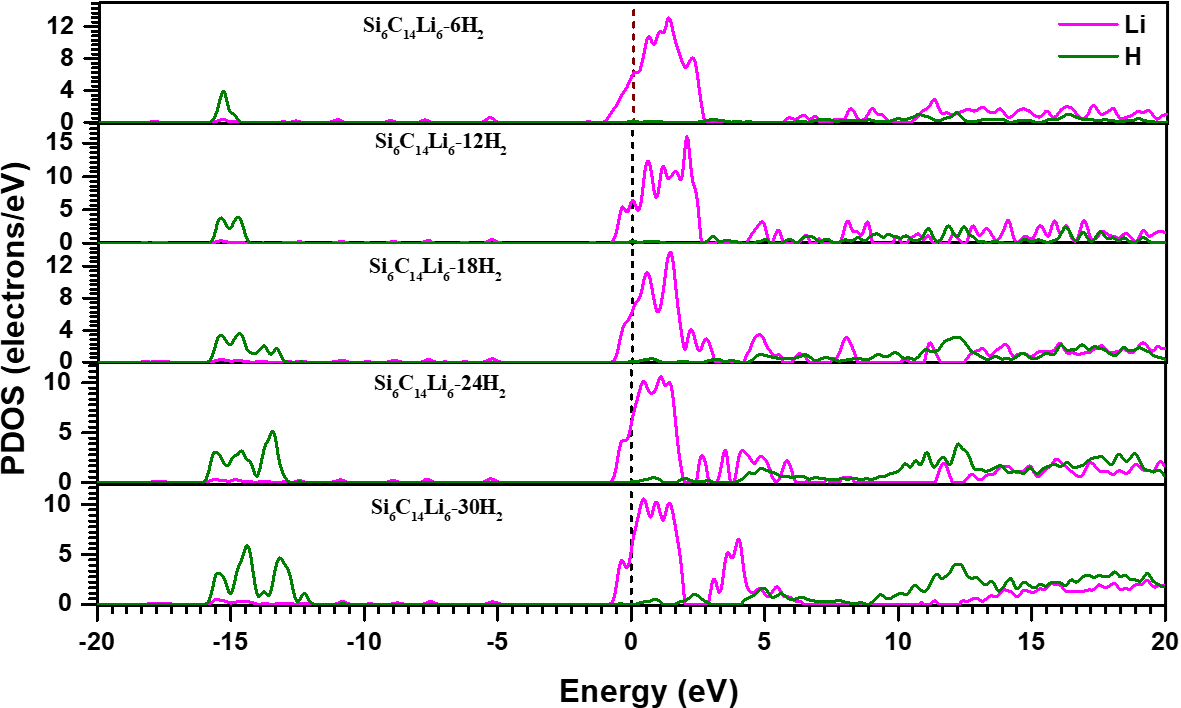}
		\caption{}
		\label{fig:fig8a}
	\end{subfigure}
	\quad
	\begin{subfigure}[h!]{0.99\textwidth}
		\centering
		\includegraphics[width=\textwidth]{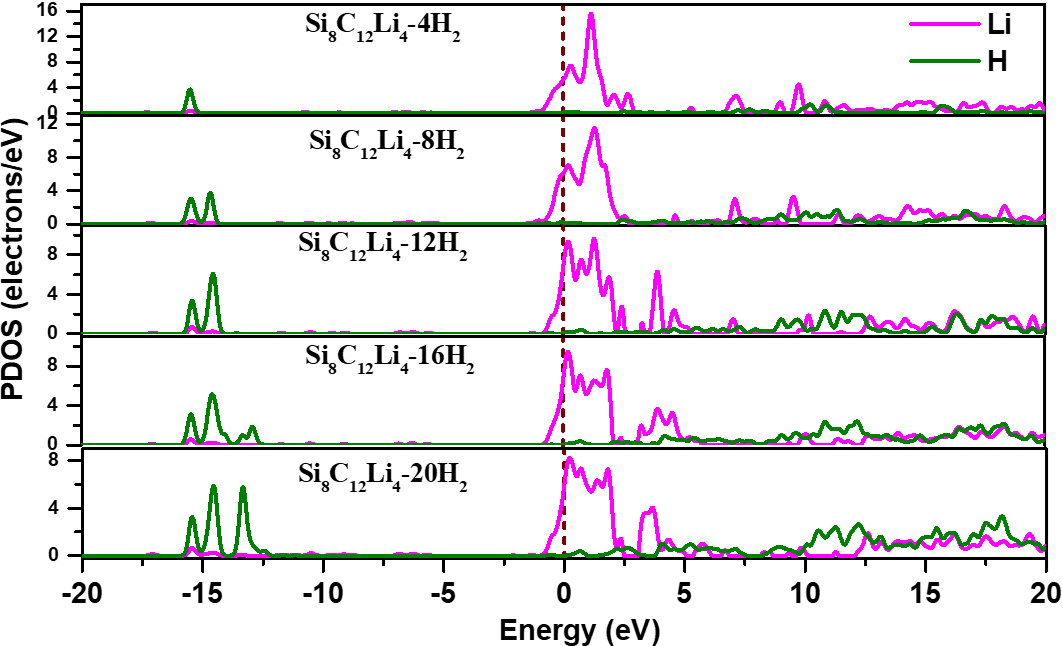}
		\caption{}
		\label{fig:fig8b}
	\end{subfigure}
	\caption{PDOS for sequential hydrogen adsorbed (a) Si$_{6}$C$_{14}$Li$_{6}$
		and (b) Si$_{8}$C$_{12}$Li$_{4}$ systems.}
	\label{fig:fig8}
\end{figure}

Partial density of states (PDOS) of constituent elements Li and H in Si$_{6}$C$_{14}$Li$_{6}$-nH$_{2}$ and Si$_{8}$C$_{12}$Li$_{4}$-nH$_{2}$ complexes are plotted using convolution of spectra to get insight into the binding mechanism and orbital hybridization. The graph for the PDOS of Si$_{6}$C$_{14}$Li$_{6}$-nH$_{2}$ and Si$_{8}$C$_{12}$Li$_{4}$-nH$_{2}$ are depicted in Figure \ref{fig:fig8}. Figure \ref{fig:fig8a} \& Figure \ref{fig:fig8b} demonstrates that the LUMO level and the other unoccupied molecular orbitals have major contribution in bonding. After H$_{2}$ adsorption over Li centers of the Si$_{6}$C$_{14}$Li$_{6}$ and Si$_{8}$C$_{12}$Li$_{4}$ nanocages, the density of states contribution of Li and H are found to undergo a significant change. With sequential adsorption of H$_{2}$ molecules over Si$_{6}$C$_{14}$Li$_{6}$ nanocage, the density of states of Li goes on decreasing whereas the density of states of the adsorbed H$_{2}$ molecules goes on increasing. As previously demonstrated by ESP maps and Hirshfed charge analyses, the reduction in density of Li states is linked with charge loss. However, the increase in density of states of the adsorbed hydrogen is associated to charge gain. Initially, the density of state peak of bare H$_{2}$ at -15 eV is sharply oriented. But due to multiple H$_{2}$ adsorption, the density of states distribution widens over the energy axis with a few short peaks which signifies weakening of interaction with successive adsorption. In addition, the degeneration and shifting of peaks of H in adsorbed H$_{2}$ molecules Si$_{6}$C$_{14}$Li$_{6}$-30H$_{2}$ denotes the charge polarization effect between Li centers and molecular hydrogen, resulting to efficient H$_{2}$ adsorption \cite{key-52}. In addition, partial overlapping of the electronic states of adsorbed hydrogen and Li in Si$_{6}$C$_{14}$Li$_{6}$-30H$_{2}$ indicate absence of strong orbital overlap between them, which confirms that H$_{2}$ molecules are held via weak Van der Waals interaction. A similar trend for variation of electronic states of the decorated Li atom and the adsorbed H$_{2}$ molecules over Si$_{8}$C$_{12}$Li$_{4}$ nanocages is observed.

\subsubsection{Topological analysis}

In principle, the macroscopic behavior of materials is determined by the interactions between constituent atoms at the microscopic level. Therefore, we conducted Bader's topological studies \cite{key-46} to study the type of interaction between the Si$_{6}$C$_{14}$Li$_{6}$ and Si$_{8}$C$_{12}$Li$_{4}$ nanocages and the adsorbed H$_{2}$ molecules over them. The interpretation of electron density ($\rho(r)$), Laplacian of electron density ($\nabla^{2}\rho$) and total energy density $H(r)$ at various bond critical points (BCPs) reveals the nature of atomic interactions. The $\rho(r)$, $\nabla^{2}\rho(r)$, $H(r)$ and bond degree (BD) at various BCPs are tabulated in the
Table \ref{tab:tab4}. The negative value of $\nabla^{2}\rho(r)$ at a given BCP implies electrons sharing, and hence it characterizes a covalent type of interaction. A relative depletion of electrons along a given bond path, as indicated by a positive $\nabla^{2}\rho(r)$, demonstrates a closed-shell type of noncovalent interactions \cite{key-55}.
In the context of weak interactions, it is necessary to consider not only the variation in electron density $\rho(r)$ and its Laplacian $\nabla^{2}\rho(r)$ at different BCPs, but also the variation in total local energy density $H(r)$. The total local energy density is defined as the sum of the local potential energy density $V(r)$ and the local kinetic energy $G(r)$ at the corresponding BCP. At a given BCP, negative values of $H(r)$ indicates bonds formed by sharing of electrons and thus contains covalent component. Positive $H(r)$ values at a certain BCP signifies closed shell type of interaction \cite{key-56}. Additionally, the characterization of interaction may be comprehended by considering bond degree (BD), which is defined as the overall local energy density per electron at a specific BCP. Mathematically, it can be represented as $BD = H/\rho{(r)}$.

\begin{table}
	\centering
	\caption{Topological parameters $\rho(r)$, $\nabla^{2}\rho(r)$ and $H(r)$
		at all BCPs of the systems.}
	\label{tab:tab4}
	\begin{tabular}{cccccc}
		\toprule
		Species & BCP & $\rho(r)$ (a.u) & $\nabla^{2}\rho(r)$ & $H(r)$ (a.u) & Bond degree (BD)\\
		\midrule 
		H$_{2}$ & H-H & 0.257316  & -1.01865  & -0.254663  & -1\\
		\hline 
		\multirow{2}{*}{Si$_{6}$C$_{14}$} & C-C & 0.277374  & -0.668508  & -0.258539  & -0.9\\
		& C-Si & 0.116673  & 0.263295  & -0.070425  & -0.6 \\
		\hline 
		\multirow{3}{*}{Si$_{6}$C$_{14}$Li$_{6}$} & C-C & 0.247713  & -0.513906  & -0.20788  & -0.8 \\
		& C-Si & 0.084015 & 0.122324  & -0.046665  & -0.6 \\
		& C-Li & 0.022687  & 0.115333  & 0.003342  & 0.1 \\
		\hline 
		\multirow{5}{*}{Si$_{6}$C$_{14}$Li$_{6}$-30H$_{2}$} & C-C & 0.248382  & -0.519464  & -0.208909  & -0.8 \\
		& C-Si & 0.081904  & 0.105188  & -0.045559  & -0.6 \\
		& C-Li & 0.02356  & 0.131348  & 0.003836  & 0.2 \\
		& Li-H$_{2}$ & 0.009148  & 0.051375  & 0.002563  & 0.3 \\
		& H-H & 0.260958  & -1.063504  & -0.266436  & -1 \\
		\hline 
		\multirow{2}{*}{Si$_{8}$C$_{12}$} & C-C & 0.353642  & -1.018858  & -0.426445  & -1.2 \\
		& C-Si & 0.100182  & 0.198428  & -0.057756  & -0.6 \\
		\hline 
		\multirow{3}{*}{Si$_{8}$C$_{12}$Li$_{4}$} & C-C & 0.319298  & -0.838761  & -0.348065  & -1.1 \\
		& C-Si & 0.106616  & 0.19604  & -0.063803  & -0.6 \\
		& C-Li & 0.024209  & 0.141691  & 0.004636  & 0.2 \\
		\hline 
		\multirow{5}{*}{Si$_{8}$C$_{12}$Li$_{4}$-20H$_{2}$} & C-C & 0.319649  & -0.840835  & -0.348774  & -1.1 \\
		& C-Si & 0.106301  & 0.1945  & -0.063546  & -0.6 \\
		& C-Li & 0.022379  & 0.127829  & 0.004346  & 0.2 \\
		& Li-H$_{2}$ & 0.008194  & 0.044993  & 0.002215  & 0.3 \\
		& H-H & 0.26193  & -1.071127  & -0.268151  & -1 \\
		\bottomrule
	\end{tabular}
\end{table}

Positive BD, also known as softening degree (SD), refers to a non-covalent interaction of the closed shell type. On the other hand, the covalent degree (CD), which is a negative BD, represents covalent interactions that involve shared and intermediate types. The stronger the covalent interaction will be, the greater the magnitude of CD \cite{key-57}. The tabulated data in Table \ref{tab:tab4} reveals that bonding nature in the hosts hardly undergoes any change during Li functionalization and successive H$_{2}$ adsorption. For the saturated hydrogen adsorbed Si$_{6}$C$_{14}$Li$_{6}$ nanocage, the $\rho(r)$ value over the H-H bond in the adsorbed H$_{2}$ molecule is 0.26 a.u, which is nearly same as  the $\rho(r)$ value in the isolated H$_{2}$ molecule. This clearly demonstrates that H$_{2}$ molecules are molecularly adsorbed onto the designed nanocages, which is in agreement with the results reported by Jiguang et al. \cite{key-58}. The sign of $\nabla^{2}\rho(r)$ and $H(r)$ in C-C and C-Si are the same in Si$_{6}$C$_{14}$, Si$_{6}$C$_{14}$Li$_{6}$, and saturated hydrogen adsorbed Si$_{6}$C$_{14}$Li$_{6}$ systems, implying that the nature of bonding hardly undergoes any significant change due to Li functionalization and successive H$_{2}$ adsorption. The positive signatures of $\nabla^{2}\rho(r)$ and $H(r)$ at the BCPs of Li-H$_{2}$ in the hydrogenated systems demonstrate that H$_{2}$ molecules are adsorbed via weak non-covalent interaction. In addition, the negative sign of $\nabla^{2}\rho(r)$ and $H(r)$ suggests that the H-H bond remains intact following the adsorption process. A Similar trend is observed in case of Si$_{8}$C$_{12}$ cage. 

\subsection{MD Simulation}

\begin{figure}[h!]
	\centering
	\begin{subfigure}[h!]{0.485\textwidth}
		\centering
		\includegraphics[width=\textwidth]{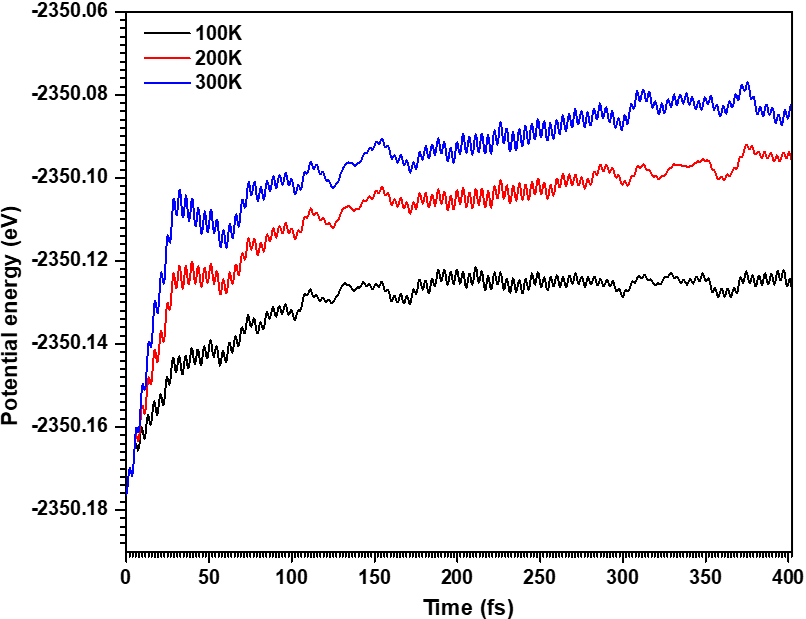}
		\caption{Si$_{6}$C$_{14}$Li$_{6}$-30H$_{2}$}
		\label{fig:fig9a}
	\end{subfigure}
	\hfill
	\begin{subfigure}[h!]{0.485\textwidth}
		\centering
		\includegraphics[width=\textwidth]{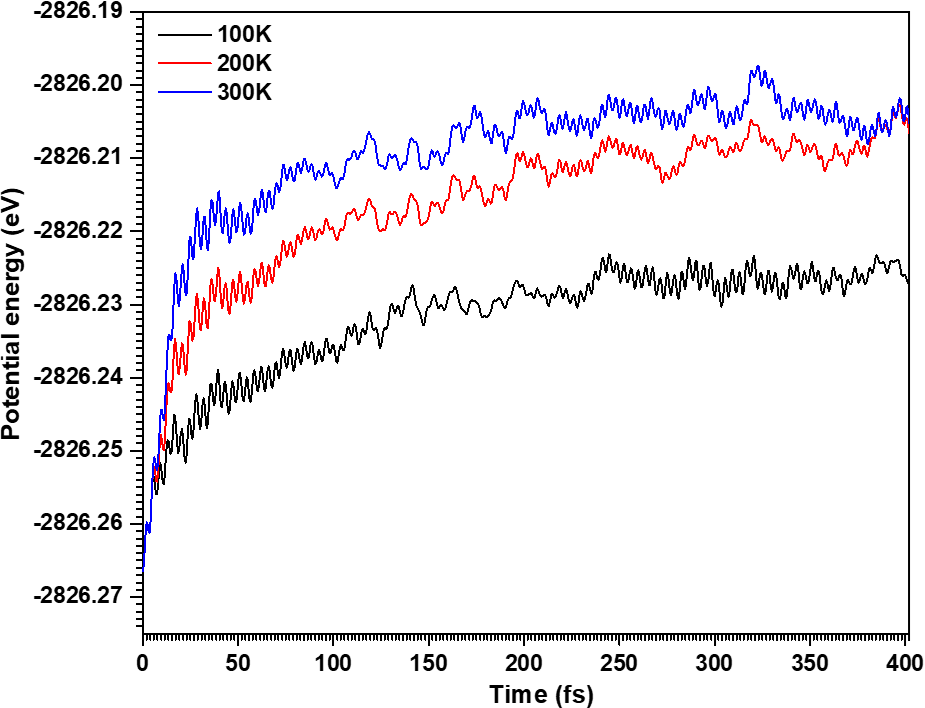}
		\caption{Si$_{8}$C$_{12}$Li$_{4}$-20H$_{2}$}
		\label{fig:fig9b}
	\end{subfigure}
	\caption{Potential energy vs. time graph for the saturated hydrogen adsorbed nanocages at various temperature.}
	\label{fig:fig9}
\end{figure}

We have performed the atom centered molecular dynamics (ADMP) simulation to understand the thermal stability as well as the structural integrity of the saturated H$_{2}$ adsorbed systems. ADMP is an expanded Lagrangian method for MD that employs the Gaussian basis function and propagates the density matrix. During simulations, the temperature (kinetic energy thermostat) is maintained by the velocity scaling method. For the saturated hydrogen adsorbed Si$_{6}$C$_{14}$Li$_{6}$-30H$_{2}$ and Si$_{8}$C$_{12}$Li$_{4}$-20H$_{2}$ nanocages, ADMP simulations are performed at a temperatures of 100K, 200K and 300K at 1 bar pressure for 400fs with a time step of 0.5fs. The potential energy vs time graph for both the saturated hydrogen adsorbed systems at100K, 200K and 300K are shown in Figure \ref{fig:fig9}.

The potential energy at higher temperature is higher because of enhanced molecular movements. At 100K, about 12H$_{2}$ molecules got desorbed from the saturated hydrogen adsorbed Si$_{6}$C$_{14}$Li$_{6}$-30H$_{2}$ system. When the temperature is increased to 200K, almost 19H$_{2}$ molecules got desorbed. Further, at 300K, 25H$_{2}$ (83.3\%) molecules
get completely desorbed from the system. Similarly, for Si$_{8}$C$_{12}$Li$_{4}$-20H$_{2}$ nanocage, 12H$_{2}$ molecules are found to get desorb from the system. When the temperature is increased to 300K, complete desorption of H$_{2}$ molecules is achieved. From the Figure \ref{fig:fig9}, we find that molecular movements is higher at higher temperature, leading to easy and fast desorption. Therefore, at 1 bar atmospheric pressure, Si$_{6}$C$_{14}$Li$_{6}$ and Si$_{8}$C$_{12}$Li$_{4}$ adsorb H$_{2}$ molecules at temperatures below 100K and undergo complete desorption at room temperature. Additionally, it can be observed that the hydrogen molecules begin to shift away from the complex, and as time passes, they move away from the vicinity of the Li atoms. After complete desorption of H$_{2}$ molecules, the designed host complex endures minimal changes and remains stable. This indicates the on-board reversibility of the system at the investigated temperatures. The desorption pattern changes as the temperature varies. These results are essential for further investigation of the on-board reversible storage capacity of the designed complexes, as temperature variation causes the desorption of hydrogen molecules, thereby maintaining the stability of the host complex.

\subsection{Practical hydrogen storage}

\begin{figure}[h!]
	\centering
	\begin{subfigure}[h!]{0.485\textwidth}
		\centering
		\includegraphics[width=\textwidth]{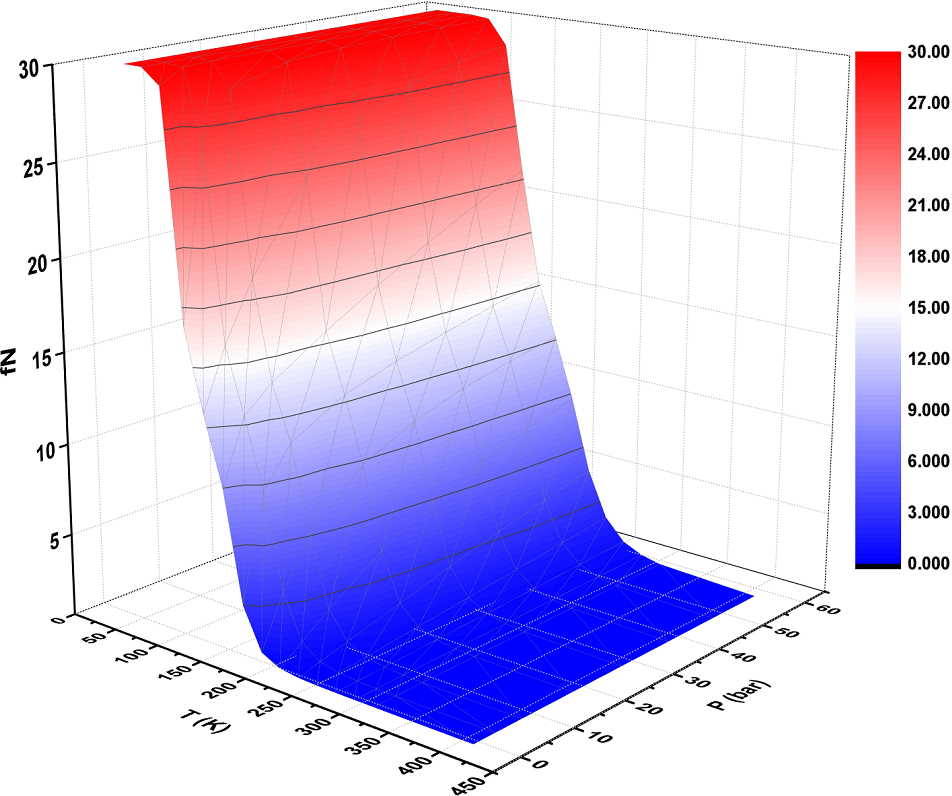}
		\caption{}
		\label{fig:fig10a}
	\end{subfigure}
	\hfill
	\begin{subfigure}[h!]{0.485\textwidth}
		\centering
		\includegraphics[width=\textwidth]{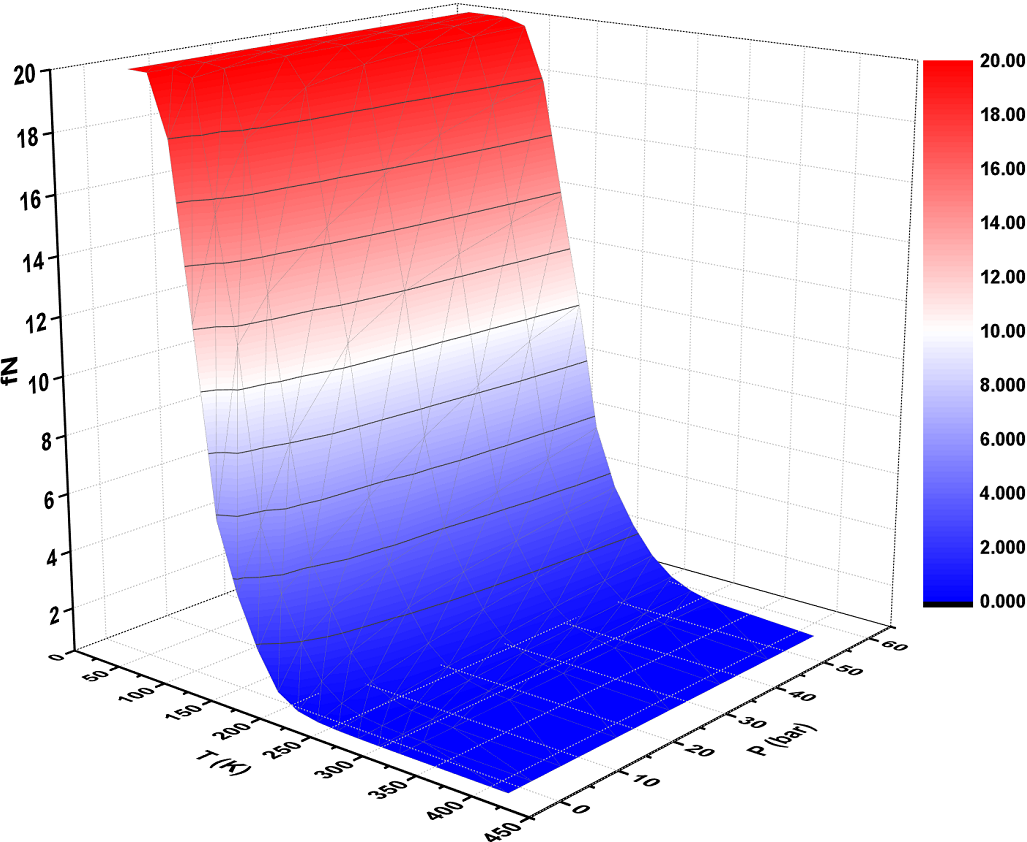}
		\caption{}
		\label{fig:fig10b}
	\end{subfigure}
	\caption{Practical hydrogen capacities of (a) Si$_{6}$C$_{14}$Li$_{6}$ and (b) Si$_{8}$C$_{12}$Li$_{4}$ nanocages at various temperature and pressure.}
	\label{fig:fig10}
\end{figure}

At a given temperature and pressure, the maximum number of usable H$_{2}$ molecules that remain adsorbed in a storage system marks the practical storage capacity of a hydrogen storage material. The practical H$_{2}$ storage capacity of the designed Si$_{6}$C$_{14}$Li$_{6}$ and Si$_{8}$C$_{12}$Li$_{4}$ nanocages clusters are investigated thoroughly using the concept of grand canonical ensemble as explained in the theoretical section. The phonon contribution in the entropy is neglected and hence the degeneracy g$_{n}$ value is taken as 1. The occupation number ($f$) denotes the number of usable H$_{2}$ molecules remain adsorbed over each sorption center of the Si$_{6}$C$_{14}$Li$_{6}$ and Si$_{8}$C$_{12}$Li$_{4}$ nanocages at a given temperature and pressure. Consequently, for $N$ sorption sites, the total number of usable H$_{2}$ molecules can be given by $fN$. Here, in our study, $N=4,6$ for Si$_{6}$C$_{14}$Li$_{6}$ and Si$_{8}$C$_{12}$Li$_{4}$ nanocages respectively. The variation of fN for the saturated hydrogen adsorbed Si$_{6}$C$_{14}$Li$_{6}$ and Si$_{8}$C$_{12}$Li$_{4}$ nanocages with temperature (T) and pressure (P) is shown in Figure \ref{fig:fig10}. In case of Si$_{6}$C$_{14}$Li$_{6}$-30H$_{2}$ (Ref. Figure \ref{fig:fig10}), we can observe that for the temperature lying in the range of 40-80K all 30H$_{2}$ molecules remain adsorbed at all pressure range of 1-60bar. When the temperature is increased to 100K at 1 bar pressure, only 17 H$_{2}$ molecules are found to remain adsorb over the system. However, at the same temperature pressure is increased, occupation number increases and in the pressure range
of 20-60 bar, almost all 30H$_{2}$ molecules are found remain adsorb over the system. Beyond this temperature, high occupation number of H$_{2}$ molecules is achieved only at higher pressure range of 40-60 bar. At 180K/60 bar, only 12 H$_{2}$ molecules are found to remain adsorbed over the temperature. Further increase in temperature allows
more H$_{2}$ molecules to get desorbed from the system. A similar trend of desorption is also followed in case of saturated hydrogen adsorbed Si$_{8}$C$_{12}$Li$_{4}$ nanocage as shown in the Figure \ref{fig:fig10b}. After thorough analysis of the adsorption-desorption pattern at various temperature and pressure for both the designed hosts, it can be estimated that 100K/60bar and 240K/1bar can be set as adsorption and desorption condition respectively. The practical gravimetric densities for the saturated hydrogen adsorbed Si$_{6}$C$_{14}$Li$_{6}$ and Si$_{8}$C$_{12}$Li$_{4}$ nanocages at the set adsorption-desorption conditions are found to be 13.73 wt\% and 9.08 wt\%.

\begin{table}[h!]
	\centering
	\caption{Practical hydrogen storage capacity obtained by setting 100K/60bar and 240K/1bar as adsorption and desorption condition respectively. N$_{ads}$ and N$_{des}$ are the occupation number of H$_{2}$ molecules at 100K/60bar and 240K/1bar. N$_{use}$ indicates number of usable hydrogen molecules and is calculated by the difference of N$_{ads}$ and N$_{des}$}
	\label{Table-prac_storage}
	\begin{tabular}{ccccccc}
		\toprule 
		Complex & $N_{th}$ & $N_{ads}$ & $N_{des}$ & $N_{use}$ & $G_{th}(wt\%)$ & $G_{prac}(wt\%)$\\
		\midrule 
		Si$_{6}$C$_{14}$Li$_{6}$-30H$_{2}$ & 30 & 29.9 & 0.03 & 29.87 & 13.78 & 13.73 \\
		Si$_{8}$C$_{12}$Li$_{4}$-20H$_{2}$ & 20 & 19.7 & 0.03 & 19.67 & 9.22 & 9.08 \\
		\bottomrule
	\end{tabular}	
\end{table}
\section{Conclusions}

In conclusion, we proposed Li decorated Si$_{6}$C$_{14}$ and Si$_{8}$C$_{12}$ nanocages as an effective hydrogen storage
media using density functional theory. The Li atoms were found to get adsorb over the pentagonal ring with one Si atom and tetragonal rings of Si$_{6}$C$_{14}$ and Si$_{8}$C$_{12}$ nanocages respectively. Global reactivity descriptors and MD simulation confirm the structural stability and integrity of the designed complexes without any clustering.
Each Li atom decorated over the designed Si$_{6}$C$_{14}$ and Si$_{8}$C$_{12}$ nanocages can adsorb about 5H$_{2}$ molecules with optimum adsorption energy, thereby achieving a gravimetric density of 13.8\% and 9.2\% respectively. The gravimetric density obtained is well above the target set by US-DoE. The major cause of multiple H$_{2}$ adsorption over
Si$_{6}$C$_{14}$Li$_{6}$ and Si$_{8}$C$_{12}$Li$_{4}$ nanocages are found to be due to charge transfer and orbital hybridization as confirmed by ESP maps and PDOS analyses. The nature of interaction between the decorated Li centers and the adsorbed H$_{2}$ molecules are found to be weak noncovalent type, as confirmed by the Badar's topological analysis. Hence, the computationally designed Si$_{6}$C$_{14}$Li$_{6}$ and Si$_{8}$C$_{12}$Li$_{4}$ nanocages can be regarded as promising template for hydrogen storage materials.

\section*{Acknowledgements}

We acknowledge our institute Indian Institute of Technology (Indian School of Mines) for providing research facilities.


\begin{thebibliography}{10}
	\bibitem{key-1}Shafiee, Shahriar, and Erkan Topal. " When
	will fossil fuel reserves be diminished?." Energy policy
	37.1 (2009): 181-189.
	
	\bibitem{key-2}Johnsson, F., Kjärstad, J., \& Rootzén, J. (2019).
	The threat to climate change mitigation posed by the abundance of
	fossil fuels. Climate Policy, 19(2), 258-274.
	
	\bibitem{key-3}Züttel, A., Remhof, A., Borgschulte, A., \& Friedrichs,
	O. (2010). Hydrogen: the future energy carrier. Philosophical Transactions
	of the Royal Society A: Mathematical, Physical and Engineering Sciences,
	368(1923), 3329-3342.
	
	\bibitem{key-4}Singh, S., Jain, S., Venkateswaran, P. S., Tiwari,
	A. K., Nouni, M. R., Pandey, J. K., \& Goel, S. (2015). Hydrogen:
	A sustainable fuel for future of the transport sector. Renewable and
	sustainable energy reviews, 51, 623-633.
	
	\bibitem{key-5}Schlapbach, L., \& Züttel, A. (2011). Hydrogen-storage
	materials for mobile applications. Materials for sustainable energy:
	a collection of peer-reviewed research and review articles from nature
	publishing group, 265-270.
	
	\bibitem{key-6}Jena, P. (2011). Materials for hydrogen storage: past,
	present, and future. The Journal of Physical Chemistry Letters, 2(3),
	206-211.
	
	\bibitem{key-7}U.S. Department of Energy's Energy Efficiency and
	Renewable Energy Website. https://www1.eere.energy.gov/hydrogen-and
	fuelcells/storage/current\_technology.html (2010).
	
	\bibitem{key-8}Züttel, A. (2003). Materials for hydrogen storage.
	Materials today, 6(9), 24-33.
	
	\bibitem{key-9}Bhatia, S. K., \& Myers, A. L. (2006). Optimum conditions
	for adsorptive storage. Langmuir, 22(4), 1688-1700.
	
	\bibitem{key-10}Mehboob, M. Y., Hussain, R., Younas, F., Jamil, S.,
	Iqbal, M. M. A., Ayub, K., ... \& Janjua, M. R. S. A. (2023). Computation
	assisted design and prediction of alkali-metal-centered B12N12 nanoclusters
	for efficient H2 adsorption: new hydrogen storage materials. Journal
	of Cluster Science, 34(3), 1237-1247.
	
	\bibitem{key-11}Jaiswal, A., Sahoo, R. K., Ray, S. S., \& Sahu, S.
	(2022). Alkali metals decorated silicon clusters (SinMn, n= 6, 10;
	M= Li, Na) as potential hydrogen storage materials: A DFT study. International
	Journal of Hydrogen Energy, 47(3), 1775-1789.
	
	\bibitem{key-12}Dong, H., Hou, T., Lee, S. T., \& Li, Y. (2015).
	New Ti-decorated B40 fullerene as a promising hydrogen storage material.
	Scientific reports, 5(1), 9952.
	
	\bibitem{key-13}Jaiswal, A., Chakraborty, B., \& Sahu, S. (2023).
	A computational insight on the effect of encapsulation and Li functionalization
	on Si12C12 heterofullerene for H2 adsorption: a strategy for effective
	hydrogen storage. ACS Applied Energy Materials, 6(6), 3374-3389.
	
	\bibitem{key-14}Ahangari, M. G., \& Mashhadzadeh, A. H. (2020). Density
	functional theory based molecular dynamics study on hydrogen storage
	capacity of C24, B12N12, Al12 N12, Be12O12, Mg12O12, and Zn12O12 nanocages.
	International Journal of Hydrogen Energy, 45(11), 6745-6756.
	
	\bibitem{key-15}Guo, L., Li, Y., Ma, Y., Liu, Y., Peng, D., Zhang,
	L., \& Han, S. (2017). Enhanced hydrogen storage capacity and reversibility
	of LiBH4 encapsulated in carbon nanocages. International Journal of
	Hydrogen Energy, 42(4), 2215-2222.
	
	\bibitem{key-16}Schneemann, A., White, J. L., Kang, S., Jeong, S.,
	Wan, L. F., Cho, E. S., ... \& Stavila, V. (2018). Nanostructured
	metal hydrides for hydrogen storage. Chemical reviews, 118(22), 10775-10839.
	
	\bibitem{key-17}Orimo, S. I., Nakamori, Y., Eliseo, J. R., Züttel,
	A., \& Jensen, C. M. (2007). Complex hydrides for hydrogen storage.
	Chemical reviews, 107(10), 4111-4132.
	
	\bibitem{key-18}Langmi, H. W., Ren, J., North, B., Mathe, M., \&
	Bessarabov, D. (2014). Hydrogen storage in metal-organic frameworks:
	a review. Electrochimica Acta, 128, 368-392.
	
	\bibitem{key-19}Froudakis, G. E. (2011). Hydrogen storage in nanotubes
	\& nanostructures. Materials Today, 14(7-8), 324-328.
	
	\bibitem{key-20}Jain, V., \& Kandasubramanian, B. (2020). Functionalized
	graphene materials for hydrogen storage. Journal of Materials Science,
	55(5), 1865-1903.
	
	\bibitem{key-21}Basic Research Needs for the Hydrogen Economy. www.sc.doe.
	gov/bes/hydrogen.pdf (2004).
	
	\bibitem{key-22}Chen, G., Gong, X. G., \& Chan, C. T. (2005). Theoretical
	study of the adsorption of H 2 on (3, 3) carbon nanotubes. Physical
	Review B, 72(4), 045444.
	
	\bibitem{key-23}Okamoto, Y., \& Miyamoto, Y. (2001). Ab initio investigation
	of physisorption of molecular hydrogen on planar and curved graphenes.
	The journal of physical chemistry B, 105(17), 3470-3474.
	
	\bibitem{key-24}Srinivasu, K., \& Ghosh, S. K. (2011). Tuning the
	metal binding energy and hydrogen storage in alkali metal decorated
	MOF-5 through boron doping: A theoretical investigation. The Journal
	of Physical Chemistry C, 115(34), 16984-16991.
	
	\bibitem{key-25}Qiu, N. X., Zhang, C. H., \& Xue, Y. (2014). Tuning
	Hydrogen Storage in Lithium-Functionalized BC2N Sheets by Doping with
	Boron and Carbon. ChemPhysChem, 15(14), 3015-3025.
	
	\bibitem{key-26}Cho, J. H., Yang, S. J., Lee, K., \& Park, C. R.
	(2011). Si doping effect on the enhanced hydrogen storage of single
	walled carbon nanotubes and graphene. International journal of hydrogen
	energy, 36(19), 12286- 12295
	
	\bibitem{key-27}Ganji, M. D., Ahmadian, N., Goodarzi, M., \& Khorrami,
	H. A. (2011). Molecular hydrogen interacting with Si-, S-and P-doped
	C60 fullerenes and carbon nanotube. Journal of Computational and Theoretical
	Nanoscience, 8(8), 1392-1399.
	
	\bibitem{key-28}Ward PA, Teprovich JA Jr, Compton RN, Schwartz V,
	Veith GM, Zidan R. Evaluation of the physi-and chemisorption of hydrogen
	in alkali (Na, Li) doped fullerenes. Int J Hydrogen Energy. 2015;40(6):2710-2716.
	
	\bibitem{key-29}Ataca, C., Aktürk, E., Ciraci, S. A. L. \.{I}. M.,
	\& Ustunel, H. A. N. D. E. (2008). High-capacity hydrogen storage
	by metallized graphene. Applied Physics Letters, 93(4), 043123.
	
	\bibitem{key-30}Ammar HY, Badran HM. Ti deposited C20 and Si20 fullerenes
	for hydrogen storage application, DFT study. Int J Hydrogen Energy.
	2021;46(27):14565-14580.
	
	\bibitem{key-31}Sun Q, Wang Q, Jena P, Kawazoe Y. Clustering of Ti
	on a C60 surface and its effect on hydrogen storage. J Am Chem Soc.
	2005;127(42):14582-14583.
	
	\bibitem{key-32}Krasnov PO, Ding F, Singh AK, Yakobson BI. Clustering
	of Sc on SWNT and reduction of hydrogen uptake: ab-initio allelectron
	calculations. J Phys Chem C. 2007;111
	
	\bibitem{key-33}Huang, J., Zhou, C., \& Duan, X. (2021). Li decorated
	C9N4 monolayer as a potential material for hydrogen storage. International
	Journal of Hydrogen Energy, 46(65), 32929-32935.
	
	\bibitem{key-34}Guo, Y., Jiang, K., Xu, B., Xia, Y., Yin, J., \&
	Liu, Z. (2012). Remarkable hydrogen storage capacity in Li-decorated
	graphyne: theoretical predication. The Journal of Physical Chemistry
	C, 116(26), 13837-13841.
	
	\bibitem{key-35}Sun Q, Jena P, Wang Q, Marquez M. First-principles
	study of hydrogen storage on Li12C60. J Am Chem Soc. 2006;128(30):
	9741-9745.
	
	\bibitem{key-36}Wang Q, Jena P. Density functional theory study of
	the interaction of hydrogen with Li6C60. J Phys Chem Lett. 2012;3(9):
	1084-1088.
	
	\bibitem{key-37}Sahoo RK, Chakraborty B, Sahu S. Reversible hydrogen
	storage on alkali metal (Li and Na) decorated C20 fullerene: a density
	functional study. Int J Hydrogen Energy. 2021;46(80):40251- 40261.
	
	\bibitem{key-38}Metin T, Parlak C, Alver Ö, Tepe M. Hydrogen storage
	and electronic properties of C20, C15M5 and H2@ C15M5 (M = Al, Si,
	Ga, Ge) nanoclusters. J Mol Struct. 2022;1247:131272.
	
	\bibitem{key-39}Huang W, Shi M, Song H, et al. Hydrogen storage on
	chainsterminated fullerene C20 with density functional theory. Chem
	Phys Lett. 2020;758:137940
	
	\bibitem{key-40}Jaiswal, A., Chakraborty, B., \& Sahu, S. (2022).
	High capacity reversible hydrogen storage in Si substituted and Li
	decorated C20 fullerene: Acumen from density functional theory simulations.
	International Journal of Energy Research, 46(14), 19521-19537.
	
	\bibitem{key-41}Koohi, M., Amiri, S. S., \& Shariati, M. (2017).
	Silicon impacts on structure, stability and aromaticity of C20-nSin
	heterofullerenes (n= 1--10): A density functional perspective. Journal
	of Molecular Structure, 1127, 522-531.
	
	\bibitem{key-42}Chattaraj, P. K., Nath, S., \& Maiti, B. (2003).
	Reactivity descriptors (pp. 295-322). Marcel Dekker: New York.
	
	\bibitem{key-43}Geerlings, P., De Proft, F., \& Langenaeker, W. (2003).
	Conceptual density functional theory. Chemical reviews, 103(5), 1793-1874.
	
	\bibitem{key-44}Koopmans, T. (1934). Über die Zuordnung von Wellenfunktionen
	und Eigenwerten zu den einzelnen Elektronen eines Atoms. physica,
	1(1-6), 104-113.
	
	\bibitem{key-45}Boys, S. F., \& Bernardi, F. J. M. P. (1970). The
	calculation of small molecular interactions by the differences of
	separate total energies. Some procedures with reduced errors. Molecular
	Physics, 19(4), 553-566. 
	
	\bibitem{key-46}Combinatorial search for optimal hydrogen-storage
	nanomaterials based on polymers. Physical review letters, 97(5), 056104
	
	\bibitem{key-47}Lee, H., Choi, W. I., Nguyen, M. C., Cha, M. H.,
	Moon, E., \& Ihm, J. (2007). Ab initio study of dihydrogen binding
	in metal-decorated polyacetylene for hydrogen storage. Physical Review
	B, 76(19), 195110.
	
	\bibitem{key-48}Lide, D. R. (Ed.). (2004). CRC handbook of chemistry
	and physics (Vol. 85). CRC press.
	
	\bibitem{key-49}Gaussian 09, Revision E.01, M. J. Frisch, G. W. Trucks,
	H. B. Schlegel, G. E. Scuseria, M. A. Robb, J. R. Cheeseman, G. Scalmani,
	V. Barone, G. A. Petersson, H. Nakatsuji, X. Li, M. Caricato, A. Marenich,
	J. Bloino, B. G. Janesko, R. Gomperts, B. Mennucci, H. P. Hratchian,
	J. V. Ortiz, A. F. Izmaylov, J. L. Sonnenberg, D. Williams-Young,
	F. Ding, F. Lipparini, F. Egidi, J. Goings, B. Peng, A. Petrone, T.
	Henderson, D. Ranasinghe, V. G. Zakrzewski, J. Gao, N. Rega, G. Zheng,
	W. Liang, M. Hada, M. Ehara, K. Toyota, R. Fukuda, J. Hasegawa, M.
	Ishida, T. Nakajima, Y. Honda, O. Kitao, H. Nakai, T. Vreven, K. Throssell,
	J. A. Montgomery, Jr., J. E. Peralta, F. Ogliaro, M. Bearpark, J.
	J. Heyd, E. Brothers, K. N. Kudin, V. N. Staroverov, T. Keith, R.
	Kobayashi, J. Normand, K. Raghavachari, A. Rendell, J. C. Burant,
	S. S. Iyengar, J. Tomasi, M. Cossi, J. M. Millam, M. Klene, C. Adamo,
	R. Cammi, J. W. Ochterski, R. L. Martin, K. Morokuma, O. Farkas, J.
	B. Foresman, and D. J. Fox, Gaussian, Inc., Wallingford CT, 2016. 
	
	\bibitem{key-50}Szabo, A., \& Ostlund, N. S. (2012). Modern quantum
	chemistry: introduction to advanced electronic structure theory. Courier
	Corporation.
	
	\bibitem{key-51}Zhao, Y., \& Truhlar, D. G. (2008). The M06 suite
	of density functionals for main group thermochemistry, thermochemical
	kinetics, noncovalent interactions, excited states, and transition
	elements: two new functionals and systematic testing of four M06-class
	functionals and 12 other functionals. Theoretical chemistry accounts,
	120(1), 215-241.Lee, H., Choi, W. I., \& Ihm, J. (2006)
	
	\bibitem{key-52}Tenderholt, A. L., \& Langner, K. M. (2008). Cclib:
	A library for package-independent computational chemistry algorithms.
	Journal of Computational Chemistry, 29(5), 839-845.
	
	\bibitem{key-53}Bader, R. F. (1991). A quantum theory of molecular
	structure and its applications. Chemical Reviews, 91(5), 893-928.Keith
	TA.
	
	\bibitem{key-54}AIMALL (version 11.02.27, standard). Overland Park,KS,
	USA: TK Gristmill Software; 2011. 
	
	\bibitem{key-55}B. Schaefer, S. Alireza Ghasemi, S. Roy, S. Goedecker,
	Stabilized quasinewton optimization of noisy potential energy surfaces,
	The Journal of chemical physics 142 (3) (2015) 034112.
	
	\bibitem{key-56}S. Goedecker, W. Hellmann, T. Lenosky, Global minimum
	determination of the born-oppenheimer surface within density functional
	theory, Physical review letters 95 (5) (2005) 055
	
	\bibitem{key-57}Korabel'nikov, D. V., \& Zhuravlev, Y. N. (2019). The nature of the chemical bond in oxyanionic crystals based on QTAIM topological analysis of electron densities. RSC advances, 9(21), 12020-12033.
	
	\bibitem{key-58}Du, J., Sun, X., Zhang, L., Zhang, C., \& Jiang, G. (2018). Hydrogen storage of Li4\&B36 cluster. Scientific Reports, 8(1), 1940.
	
	\bibitem{key-59}Niu, J., Rao, B. K., \& Jena, P. (1992). Binding
	of hydrogen molecules by a transition-metal ion. Physical review letters,
	68(15), 2277..
	
	\bibitem{key-60}Pearson, R. G. The principle of maximum hardness.
	Acc. Chem. Res. 1993, 26 (5), 250\textminus 255.
	
	\bibitem{key-61}Chattaraj, P. K., \& Poddar, A. (1999). Molecular
	reactivity in the ground and excited electronic states through density-dependent
	local and global reactivity parameters. The Journal of Physical Chemistry
	A, 103(43), 8691-8699. 
	
	\bibitem{key-62}Bader, R. F. W., Nguyen-Dang, T. T., \& Tal, Y. (1981).
	A topological theory of molecular structure. Reports on Progress in
	Physics, 44(8), 893.
	
	\bibitem{key-63}Espinosa, E.; Alkorta, I.; Elguero, J.; Molins, E.
	From weak to strong interactions: A comprehensive analysis of the
	topological and energetic properties of the electron density distribution
	involving X\textminus{} H···F\textminus Y systems. J. Chem. Phys.
	2002, 117 (12), 5529\textminus 5542.
	
	\bibitem{key-64}Du, J.; Sun, X.; Zhang, L.; Zhang, C.; Jiang, G.
	Hydrogen storage of Li4\&B36 cluster. Sci. Rep. 2018, 8 (1), 1\textminus 7.
	
	\bibitem{key-65}Korabel nikov, D. V.; Zhuravlev, Y. N. The nature
	of the chemical bond in oxyanionic crystals based on QTAIM topological
	analysis of electron densities. RSC Adv. 2019, 9 (21), 12020\textminus 12033.
\end{thebibliography}
\end{document}